\documentclass[11pt,a4paper]{article}

\usepackage{color}

\usepackage{booktabs}

\definecolor{gbcolor}{rgb}{.8,.1,.02}

\usepackage[table]{xcolor}
\usepackage{array}

\font\tenrsfs=rsfs10 at 12pt
\font\sevenrsfs=rsfs7
\font\fiversfs=rsfs5
\newfam\rsfsfam

\textfont\rsfsfam=\tenrsfs
\scriptfont\rsfsfam=\sevenrsfs
\scriptscriptfont\rsfsfam=\fiversfs
\def\mathscr#1{{\fam\rsfsfam\relax#1}}

\def\circa#1{\,\raise.3ex\hbox{$#1$\kern-.75em\lower1ex\hbox{$\sim$}}\,}

\usepackage{amsmath}
\usepackage{amsfonts}
\usepackage{amssymb}

\usepackage[hyperfootnotes=false]{hyperref}
\usepackage{graphicx}

\newcommand{\eq}[1]{(\ref{#1})}
\newcommand{\be}{\begin{equation}}
\newcommand{\ee}{\end{equation}}
\newcommand{\bea}{\begin{eqnarray}}
\newcommand{\ena}{\end{eqnarray}}
\newcommand{\no}{\noindent}

\newcommand{\de}{\partial}

\newcommand{\ba}{\begin{eqnarray}}
\newcommand{\ea}{\end{eqnarray}}

\makeatletter
\def\ps@mine{%
    \def\@oddfoot{\hfil\thepage\hfil}\let\@evenfoot\@oddfoot
    \let\@oddhead\@evenhead%
    \let\@mkboth\@gobbletwo
    \let\sectionmark\@gobble
    \let\subsectionmark\@gobble
    }
\renewcommand\section{\@startsection {section}{1}{\z@}%
                                   {-3.5ex \@plus -1ex \@minus -.2ex}%
                                   {2ex \@plus.2ex}%
                                   {\normalfont\large\sffamily\bfseries}}
\renewcommand\subsection{\@startsection {subsection}{1}{\z@}%
                                   {-3.5ex \@plus -1ex \@minus -.2ex}%
                                   {2ex \@plus.2ex}%
                                   {\normalfont\sffamily\bfseries}}
\makeatother

\usepackage{multirow}

\numberwithin{equation}{section}

\usepackage{cite}

\usepackage{setspace}

\allowdisplaybreaks

\begin{document}

\thispagestyle{empty}
\vspace*{-2.5cm}
\begin{minipage}{.45\linewidth}
\begin{flushleft}                           
{\footnotesize CERN-TH-2016-098}
\end{flushleft} 
\end{minipage}
\vspace{2.5cm}

\begin{center}
{\huge\sffamily\bfseries 
Thermodynamics of perfect fluids\\ \vspace{0.1cm} from scalar field theory}
 \end{center}
 
 \vspace{0.5cm}
 
 \begin{center} 
{\sffamily\bfseries \large  Guillermo Ballesteros}$^{a,b}$,   {\sffamily\bfseries \large Denis Comelli}$^c$, {\sffamily\bfseries \large Luigi Pilo$^{d,e}$}\\[2ex]
  {\it
$^a$ Institut de Physique Th\'eorique, Universit\'e Paris Saclay, CEA, CNRS\\91191 Gif-sur-Yvette, France\\\vspace{0.1cm}
$^b$CERN, Theory Division, 1211 Geneva, Switzerland\\\vspace{0.1cm}
$^c$INFN, Sezione di Ferrara,  44124 Ferrara, Italy\\\vspace{0.1cm}
$^d$Dipartimento di Scienze Fisiche e Chimiche, Universit\`a di L'Aquila,  I-67010 L'Aquila, Italy\\\vspace{0.1cm}
$^e$INFN, Laboratori Nazionali del Gran Sasso, I-67010 Assergi, Italy\\\vspace{0.3cm}
{\tt guillermo.ballesteros@cea.fr}, {\tt comelli@fe.infn.it}, {\tt luigi.pilo@aquila.infn.it}
}
\end{center}

\vspace{0.7cm}

\begin{center}
{\small \today}
\end{center}

\vspace{0.7cm}

\begin{center}
{\sc Abstract}

\end{center}
\no

The low-energy dynamics of relativistic continuous media is given by a shift-symmetric effective theory of four scalar fields. These scalars describe the embedding in spacetime of the medium and play the role of St\"uckelberg fields for spontaneously broken spatial and time translations. Perfect fluids 
are selected imposing a stronger symmetry  group or reducing the field content to a single scalar.
We explore the relation between  the field theory description of
perfect fluids to thermodynamics. By  drawing the correspondence
between the allowed operators at leading order in derivatives and the
thermodynamic variables, we find that a complete thermodynamic picture
requires the four St\"uckelberg fields.  We show that thermodynamic
stability plus the null-energy condition imply dynamical stability. 
We also argue that a consistent thermodynamic interpretation is not possible if any of the shift symmetries is explicitly broken.

\newpage

\begin{spacing}{0.90}
\setcounter{tocdepth}{1}
\tableofcontents
\end{spacing}

\section{Introduction}

Fluid dynamics and thermodynamics are
probably the oldest and better known examples of effective
descriptions of a complicated underlying system in terms of a small
number of  macroscopic variables.
Systems that admit a fluid description are found in nature at widely separate distance scales and energy regimes:
from cosmological and astrophysical applications 
to heavy-ion physics and  nonrelativistic condensed matter.
A convenient formulation of fluid dynamics in the nondissipative
limit is the {\it pull-back formalism} --see \cite{Andersson:2006nr} for a
review--, where a fluid is described through an ensemble of three derivatively coupled scalars that are interpreted as
comoving coordinates of the fluid's elements. Within this formalism, fluid dynamics can be derived from an unconstrained action
principle. A related approach was developed separately to obtain a field theory, symmetry driven,  description of the fluctuations --sound waves-- propagating in fluids and other types of continuous nonrelativistic media, see \cite{Leutwyler:1993gf,Leutwyler:1996er}. The relevant degrees of freedom from this point of view, which we can call {\it phonons}, can be identified with the Goldstone bosons of spontaneously broken translational symmetries in the pull-back formalism. Given this, the two approaches can be blended together into a fully relativistic effective field theory (EFT) of continuous media \cite{Dubovsky:2005xd,Dubovsky:2011sj}, which turns to have an ample range of applications. In order to describe continuous media beyond anisotropic elastic solids, the field content of the pull-back formalism must be extended with a fourth scalar  \cite{Dubovsky:2011sj}. This allows to include superfluids in the picture and also more complex objects that are not (yet, maybe) found in nature, such as supersolids; see also \cite{Nicolis:2015sra} for other possible types of media. The fourth scalar can be interpreted as the carrier of an extra $U(1)$ charge \cite{Dubovsky:2011sj} or as  an internal time coordinate of the self-gravitating medium \cite{Ballesteros:2016gwc}, offering  a suggestive link to massive gravity theories and, in general, models of modified gravity, see \cite{Dubovsky:2005xd,Ballesteros:2016gwc} and references therein. 

In this work we focus on perfect fluids, which correspond to two specific subclasses of the EFT of continuous media at leading order in derivatives (LO), as Figure \ref{fig:cartoon} illustrates.  Although these systems can be considered the simplest ones at the level of the energy-momentum tensor, they are not free of subtleties \cite{Dubovsky:2011sj}, and there is an ongoing effort towards understanding their properties in depth. Here we build upon the work of \cite{Dubovsky:2011sj} --see also \cite{Dubovsky:2005xd}--, where the thermodynamic interpretation of effective perfect fluids was studied. We extend the  thermodynamic correspondences proposed in \cite{Dubovsky:2005xd,Dubovsky:2011sj}, obtaining a   thermodynamic dictionary that we have condensed in Table \ref{tab:thermo}.
 
Our analysis leads to the conclusion that a general thermodynamic picture (away from specific limits) requires  indeed four scalar fields (instead of just three) and, consequently,  implies an extension of the pull-back formalism. Remarkably, the form of the action required for such a thermodynamic picture is determined by a symmetry group that constitutes a specific set of continuous field redefinitions, selecting just two effective operators \cite{Dubovsky:2011sj}. 

We  show that  a consistent thermodynamic interpretation requires, in
any case, a shift symmetry for each field in the effective
action. This is interesting because such symmetry is precisely the
minimal requirement to have an EFT organized as a derivative
expansion, see \cite{Dubovsky:2005xd, Dubovsky:2011sj,
  Nicolis:2013lma,  Ballesteros:2014sxa}.  Moreover, shift symmetries
are essential for the understanding of phonons --the degrees of
freedom responsible for the propagation of sound-- as Goldstone bosons
\cite{Leutwyler:1993gf, Leutwyler:1996er} and also \cite{Matarrese:1984zw}.

Finally, we argue that thermodynamic stability of perfect fluids plus the null-energy condition guarantee  dynamical stability, i.e.\ the absence of ghost degrees of freedom and of exponential growth of fluctuations (around  Minkowski spacetime). This holds true for all the types of perfect fluids allowed by the EFT. 

Whereas the existence of an effective action description of non-dissipative fluid dynamics should not be surprising, the fact that this action leads to a complete thermodynamic description of perfect fluids is remarkable and far reaching. Having a unified and general relativistic description odissipative dynamics and thermodynamics at the action level may open the possibility for novel applications of the pull-back formalism and the effective theory of fluids.

\section{Thermodynamics in a nutshell}
\label{sect:basics}
Thermodynamics assumes that the equilibrium states of a simple system can be entirely characterized by the extensive variables {\it volume}, $V$, {\it  energy}, $E$, and {\it particle number} of each species, $N_i$. In addition, it postulates the existence of a function of the extensive variables: the {\it entropy}, $S$, which is maximized in  the  evolution of the system. These two assumptions constitute the {\it first and second postulates of thermodynamics}. The {\it third and fourth postulates} establish, respectively, that in a composite system the entropy is an additive function over the constituent subsystems and that the entropy vanishes at zero temperature \cite{Callen}.

For our purposes, it is convenient to use intensive variables, defined by dividing the extensive variables over the volume. A simple scaling argument shows that $s=S/V$, the {\it entropy density}, is a function of the {\it energy density}, $\rho=E/V$, and the {\it particle number density}, $n=N/V$; namely\footnote{In relativistic hydrodynamics $n$ is usually meant to represent a charge density.}
\begin{align}  \label{srep}
s=s(\rho,n)\,,
\end{align}
which constitutes the {\it fundamental relation} containing all the thermodynamic information of any simple system. Expressing this relation in the equivalent {\it energy representation},
\begin{align} \label{erep}
\rho=\rho(s,n)\,,
\end{align} 
and taking its differential, we get the {\it first principle of thermodynamics}:
\begin{align}\label{termo1}
d\rho=T  \, ds+ \mu \, dn\,,
\end{align}
where the {\it temperature} and the {\it chemical potential} are defined as 
\begin{align}
 T\equiv\left.  \frac{\partial \rho}{\partial s}\right|_{n } \, , \qquad
\qquad \mu\equiv\left.  \frac{\partial \rho}{\partial n}\right|_{s }\, .
\end{align}
From the additivity of the energy $E$ and the entropy $S$, the {\it Euler relation} follows:
\begin{align}
\label{termo2}
\rho+p=T\;s+\mu\;n \, ,
\end{align}
where $p$ is the intensive variable {\it pressure}:
\begin{align}
p=\left.-\frac{\partial E}{\partial V}\right|_{S,N}\,.
\end{align}
The differential of the Euler relation, together with the first
principle, leads to the fact the intensive variables $p$, $T$ and $\mu$
are not independent but satisfy the Gibbs-Duhem relation:
\be
\label{termo3}
d p=s\,dT+n\,d\mu\,.
\ee
Given two equations among (\ref{termo1}), (\ref{termo2}) and (\ref{termo3}), the third  follows.

From the definitions of the intensive variables $p$, $T$ and $\mu$,
three equations of state can be written in the energy representation:
\be
T=T(s,\,n),\quad p=p(s,\,n),\quad \mu=\mu(s,\,n) \, .
\ee
All together, these three equations of state are equivalent to the fundamental relation \eq{erep} or \eq{srep}. Keeping only one or two of them among the three leads to some
information about the system. Clearly, the equations  of state can also be expressed in different ways depending on the two independent variables that are chosen. In general, given a simple system, two thermodynamic variables  among $\{s,\,T,\,n,\,\mu,\,\rho,\,p\}$
can be taken as independent.
  
 Starting from the energy  representation \eq{erep}, one can define other thermodynamic potentials, which are naturally associated to specific choices of pairs of independent variables different from $\{s,n\}$. For instance the Gibbs-Duhem relation \eq{termo3}  already tells us that we can use the pressure as a thermodynamic potential, which corresponds to (minus) the {\it grand potential (density)} $\omega$, i.e.\ $\omega =-p$. As \eq{termo3} indicates, $\mu$ and $T$ are the associated independent variables in this case. We can also introduce the {\it free energy density} $\mathcal{F}=\rho-T\,s$, which, using \eq{termo1} and \eq{termo2}, satisfies $d\mathcal{F}=\mu\, dn - s\,dT$, effectively selecting $n$ and $T$. Similarly, we define another potential density --which bears no standard name--, $\mathcal{I}=\rho-\mu\,n$ so $d\mathcal{I}=T\,ds-n\,d\mu$. These are all the potentials we will need for our purposes.  The relations among them are summarized in Table \ref{tab:Leg}.
\begin{table}[ht]
  \footnotesize
  \small
  \renewcommand\arraystretch{2.2}
  \centering
  \begin{tabular}{|c|c|c|c|c|}
    \hline
Thermodynamic potential  & Independent variables & Legendre
                                                      transf. to $\rho$ & Conjugate variables\\
    \hline \hline
energy density $\rho$ & $s$, $n$ & none & $T = \frac{\de \rho}{\de s}\big|_n$, $\mu =
                        \frac{\de \rho}{\de n}\big|_s$\\ \hline
free energy density ${\cal F}$ & $T$, $n$ & ${\cal F}= \rho - T \, s$
                      & $s = -\frac{\de {\cal F}}{\de T}\big|_n$, $\mu =
                        \frac{\de {\cal F}}{\de n}\big|_T$ \\ \hline
grand potential density $\omega$ & $T$, $\mu$ & $\omega=-p =\rho - T
                                                \, s - \mu \, n$ & $s =- \frac{\de \omega}{\de T}\big|_\mu$, $n =-
                        \frac{\de \omega}{\de \mu}\big|_T$\\ \hline
potential density ${\cal I}$ & $s$, $\mu$ & ${\cal I}= \rho - \mu
                                            \, n$ & $T = \frac{\de
                                                    {\cal I}}{\de s}\big|_\mu$, $n =-
                        \frac{\de {\cal I}}{\de \mu}\big|_s$\\ \hline 
\end{tabular}
  \caption{\it \small Thermodynamic potentials and  variables.  \label{tab:Leg}}
 \end{table}

\section{Energy-momentum tensor and conserved currents of perfect fluids}\label{pivo}

The energy momentum tensor (EMT), $T_{\mu\nu}$, and the currents of a macroscopically continuous system play a pivotal role in the determination of its thermodynamic interpretation (if any). Of  particular relevance are the entropy and particle currents, $s^{\mu}$ and $n^{\mu}$, respectively. In agreement with the second postulate of thermodynamics,
\be
 \nabla_\mu\,s^{\mu}\geq 0 \,,
\ee
where the equality applies only for reversible (nondissipative) processes.

In this work we focus on {\it perfect fluids}, which by the definition are the continuous media whose energy momentum tensor is of the form:  
\begin{align}
T_{\mu\nu}=(\rho+p)\;v_\mu\;v_\nu+p\;g_{\mu\nu} \,, \qquad \quad v^\mu
  v_\mu=-1 \, ,
\label{pfEMT}
\end{align}
where the four-velocity of the fluid, $v_\mu$, is the (unique)
timelike eigenvector with unit norm of $T_{\mu\nu}$~\footnote{Throughout
  the paper we use the metric signature $(-,+,+,+)$.}. 
For a perfect fluid the entropy and the particle number currents are both parallel to $v_\mu$, and so we write:
\be
\label{curr}
n_{\mu} =n\;v_{\mu},\qquad s_{\mu} =s\;v_{\mu} \,.
\ee
By using equation (\ref{termo2}), the perfect fluid EMT can be expressed as
\be \label{T4}
T_{\mu\nu}=(T \, s_\mu+\mu\, n_\mu) v_\nu+p \, g_{\mu\nu} \,.
\ee
Applying~\footnote{We thank S. Matarrese for pointing out the
  paper~\cite{Matarrese:1984zw} in which a similar treatment to
  the one in this section is given.} the conservation of the EMT to \eq{T4}, the projection $v^\mu\nabla^\nu T_{\mu\nu}=0$ leads to
\be
T\;\nabla^\alpha s_\alpha+\mu\;\nabla^\alpha n_\alpha =0 \,,
\label{ns}
\ee
where we have used the Gibbs-Duhem relation \eq{termo3} along the fluid flow: $p'= n
\, \mu' + s \, T'$, denoting by $f'=v^\mu \nabla_\mu f$ the Lie derivative of a function $f$ along the four-velocity of the fluid. The equation \eq{ns} is equivalent to the {\it first principle of thermodynamics} and it tells us that for a perfect fluid (with non-vanishing chemical potential, $\mu$) a change in the particle current implies a variation of the entropy current. 
If the particle number current is conserved (or if $\mu$ is negligible), the entropy current is also conserved and we can write $s'+\theta\, s=0$, where $\theta=\nabla^\mu v_\mu$ is the {\it expansion}. In particular,  if  $\nabla^\mu n_\mu=0$, the equation $n'+\theta\, n=0$ also holds and we find that the {\it entropy per particle} $\sigma= s/n$ is conserved along the flow lines: 
$\sigma' =0$, thus defining an {\it adiabatic fluid}. 
Instead, an {\  isentropic } fluid is defined as one that has constant entropy per particle, i.e.\ $\nabla_\mu \sigma=0$. 

Using $h^{\mu\nu}=g^{\mu\nu}+v^\mu v^\nu$ to project the conservation of the EMT orthogonally to $u_\mu$, we get the Euler equations for a perfect fluid,
\be
\label{dp}
(p+\rho) \, a_\mu +h^\nu_\mu\,\nabla_\nu\,p=0 \, , 
\ee
where $(p+\rho)/n$ is the {\it enthalpy per particle} and $a_\mu= v^\nu \nabla_\nu v_\mu $ is the {\it acceleration} along the flow lines.
This equation manifestly show that the acceleration $a_\mu$ depends on the pressure gradient, as expected, since $\rho+p$ is the relativistic generalization of the mass density.
 
Defining the vorticity tensor as, see e.g.\ \cite{rezzolla2013relativistic}:
\be
\Omega_{\mu \nu}=(\nabla_\nu\,w_\mu-\nabla_\mu\,w_\nu)  \,,
\ee
from the enthalpy current $w^\mu=(\rho+p)/n\, v^\mu$, we can express the conservation of the EMT in terms of the Carter-Lichnerowicz equations:
\be \label{CL}
n \, \Omega_{\alpha \nu}\,v^\nu= n \, T\;\nabla_\alpha \sigma-
w_\alpha \, \nabla^\nu n_\nu\, .
\ee

Indeed, (\ref{ns}) and (\ref{dp}) come
from projecting \eq{CL} along $v^\mu$ and orthogonally to it. 

A perfect fluid is sometimes said to be {\it irrotational}  if
$\Omega_{\mu \nu}=0$. In this sense, an irrotational perfect fluid
with $\nabla^\mu n_\mu=0$ satisfies $\nabla_\mu \sigma=0$ and is
called {\it isentropic}. 
Clearly, a fluid that is adiabatic ($\sigma' =0$)  is not necessarily irrotational in the previous sense.

It is also common usage to call irrotational a perfect fluid whose four-velocity is the derivative of a scalar quantity, i.e.\ $v_\mu=\partial_\mu\Psi$. Clearly, this notion is not equivalent, in general, to $\Omega_{\mu \nu}=0$.

\section{Effective action for non-dissipative hydrodynamics}
\label{sect:fluida}

The study of relativistic fluid dynamics from an unconstrained  action principle has a long history. In this work we will closely follow the treatment of the subject that we gave in \cite{Ballesteros:2016gwc}. Our approach is related to Carter's geometrical formulation of the problem \cite{Carter:1987qr}, but embedded in an effective field theory (EFT) framework; as proposed e.g.\ in \cite{Leutwyler:1993gf,Leutwyler:1996er} and later in \cite{Dubovsky:2005xd}; see also \cite{Dubovsky:2011sj}. Various aspects and applications of the EFT approach for describing continuous media have been developed in \cite{Dubovsky:2005xd,Son:2005ak, Endlich:2010hf, Dubovsky:2011sj,Nicolis:2011cs, Endlich:2012pz, Ballesteros:2012kv, Bhattacharya:2012zx, Hoyos:2012dh,  Nicolis:2013lma, Ballesteros:2013nwa, Delacretaz:2014jka, Gripaios:2014yha, Ballesteros:2014sxa, Nicolis:2015sra}. Interestingly, this framework can be used, for instance,  to treat massive and modified gravity in a unified way, interpreting them as self-gravitating media, see \cite{Ballesteros:2016gwc}.\footnote{See also \cite{ArkaniHamed:2002sp, Dubovsky:2004sg,Rubakov:2008nh} for previous works on massive gravity using the St\"uckelberg ``trick''.}  For the development of Carter's idea into the subsequent {\it pull-back formalism}, see \cite{Comer:1993zfa,Comer:1994tw}. A review of this formalism is given in \cite{Andersson:2006nr} and recent applications in the context of cosmology can be found in \cite{Comer:2011ss,
Blas:2012vn,Endlich:2012pz,Ballesteros:2012kv,
Ballesteros:2013nwa,Pourtsidou:2013nha,Ballesteros:2014sxa,Berezhiani:2015pia, Ballesteros:2016gwc,Kopp:2016mhm,Kang:2015uha}. The idea that lies at the core of this treatment consists in using the Lagrangian coordinates of a continuous medium as the low-energy degrees of freedom of the EFT. In the original pull-back formalism, three scalar fields $\Phi^a$, $a=1,2,3$, identify the fluid elements as they propagate in space. Here we add a fourth scalar $\Phi^0$, which may be interpreted at this stage as an internal time coordinate of the medium. Indeed, these scalars can be seen as St\"uckelberg fields restoring broken diffeomorphisms in four-dimensional spacetimes, see e.g.\ \cite{ArkaniHamed:2002sp, Dubovsky:2004sg,Rubakov:2008nh,Ballesteros:2016gwc}.

We require that the action of these four  fields respects the shift symmetries 
\begin{align} \label{shift}
\Phi^A\to\Phi^A+f^A\,,\quad \de_\mu f^A=0\,,\quad A,\mu=0,1,2,3\,.
\end{align}
In addition, we impose invariance under internal spatial volume preserving diffeomorphisms
\begin{align} \label{vdiff}
V_s \text{Diff}: \quad \Phi^a\to\Psi^a(\Phi^b)\,,\quad \det \left(\frac{\partial \Psi^a}{\partial{\Phi^b}}\right)=1\,,\quad a,b=1,2,3\,.
\end{align}
With these symmetries, at leading order in the derivative expansion (LO) there are only two time-like independent four-vectors and just three independent scalar operators, which can be chosen to be 
\ba \label{vels}
u^{\mu}=-\frac{\epsilon^{\mu\alpha\beta\gamma}}{6\, b\,
   \sqrt{-g}}\epsilon_{abc}\;
\partial_{\alpha}\Phi^a\;\partial_{\beta}\Phi^b\;\partial_{\gamma}\Phi^c\,,\quad
\qquad {\cal V}^{\mu}=-\frac{\partial^{\mu}\Phi^0}{\sqrt{-X}}
\ea
and 
\ba
 b=\sqrt{\det \pmb B} \, ,\quad 
 X =\partial_\mu\Phi^0\partial^\mu\Phi^0 \, , \quad 
  Y = u^\mu\;\partial_\mu\Phi^0
\,,
\ea
where the four-vectors have norm $-1$ and  $\pmb B$ is the $3\times 3$ matrix of components $B^{ab} =\partial^\mu
\Phi^a \, \partial_\mu \Phi^b$. Then, the LO action including gravity\footnote{As explained in \cite{Ballesteros:2014sxa}, the Einstein-Hilbert term is the unique possible choice at this order in derivatives.}  is \cite{Dubovsky:2011sj,Nicolis:2011cs} (see also \cite{Ballesteros:2016gwc}):
\begin{align}
S=M_{pl}^2\int d^4 x\,\sqrt{-g}\, R+\int d^4x\sqrt{-g} \, U(b,\,Y, \, X)\,,
\label{genfluid}
\end{align}
where $M_{pl}^2=1/(16\pi G)$, with $G$ being Newton's constant. The gravitational EMT tensor  is
\be
T_{\mu\nu}= \left(U - b \, U_b  \right) \,g_{\mu\nu} +\left(Y \,U_Y-b\, U_b
\right) \, u_\mu \, u_\nu + 2 X \, U_X \, {\cal V}_\mu \,  {\cal V}_\nu\,,
\label{genEMT}
\ee
where $U$ is an arbitrary {\it master function} and we have denoted by
$U_b$, $U_X$ and $U_Y$ its partial derivatives. 
Since $u^\mu$ and $\mathcal{V}^\mu$ are, in general, not parallel, this EMT does not describe a perfect fluid like (\ref{pfEMT}). Actually, \eq{genEMT} was proposed as the EMT of a superfluid; see \cite{khalatnikov1982relativistic,PhysRevD.45.4536} (and also e.g.\ \cite{Comer:1994tw,Nicolis:2011cs} within the framework we use). Indeed, since ${\cal V}_\mu\sim\partial_\mu\Phi^0$ is irrotational, it has been associated to the intrinsic
superfluid component of the medium, whereas $u_\mu$ would correspond to the standard fluid component.

In this work we will be solely interested in perfect fluids, which can be obtained either from $u^\mu$ or $\mathcal{V}^\mu$ using the action \eq{genfluid}, as the EMT \eq{genEMT} shows neatly. Concretely, if $U=U(b,Y)$ or $U=U(X)$, the resulting medium is a perfect fluid. Remarkably, the first of these two cases is obtained requiring that the action has to be invariant under~\cite{Dubovsky:2011sj} (see also e.g.\ \cite{Nicolis:2011cs,Nicolis:2013lma,Ballesteros:2016gwc}):
\begin{align} \label{ts}
T_s:\quad \Phi^0 \to \Phi^0 + f(\Phi^a)\,,\quad a=1,2,3\,,
\end{align}
with $f$ being an arbitrary function. Clearly, $U(b)$ and $U(Y)$ also
describe perfect fluids. Assuming that the action depends only on the
spatial St\"uckelbergs $\Phi^a$, $a=1,2,3$, the only possibility
respecting $V_s$Diff is $U(b)$. Similarly, $U(X)$ is the only
possibility if the action contains only $\Phi^0$ and respects a shift
symmetry. These two types of perfect fluids, $U(b)$ and $U(X)$, are
manifestly different from each other since $\mathcal{V}^\mu$ is
irrotational --in the sense that it is the gradient of the scalar $\Phi^0$-- and $u^\mu$ is not. 

At higher orders in the derivative expansion the symmetries we have discussed do not protect the perfect fluid form of the EMT of these systems, which generically acquires other terms; see e.g.\ \cite{Dubovsky:2011sj,Bhattacharya:2012zx,Ballesteros:2014sxa}. This means that from the point of view of the EFT, the ``perfectness'' of perfect fluids is only an approximate low-energy feature. 

\begin{figure}
\begin{center}\vspace{-2.7cm}
\includegraphics[width=0.60\textwidth]{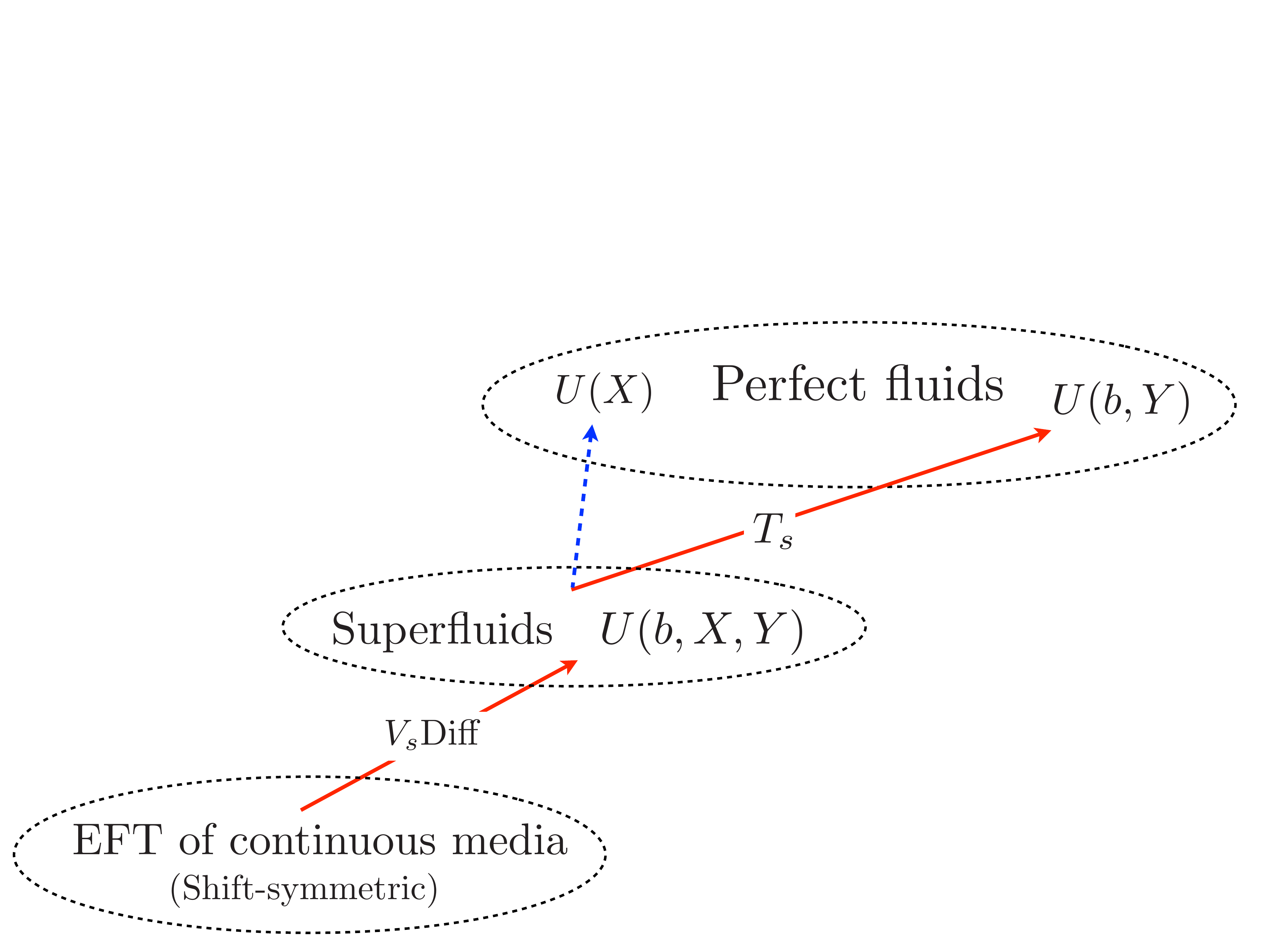}
\caption{\small \it  Red continuous arrows represent the symmetries \eq{vdiff} and \eq{ts} leading to perfect fluids at leading order in derivatives in the EFT of nondissipative continuous media.The blue dashed arrow indicates that restricting the field content to a single (temporal) St\"uckelberg field leads to (irrotational) perfect fluids.}  
\label{fig:cartoon}       
\end{center}
\end{figure}

\subsection{Conserved currents of effective perfect fluids}

The different kinds of perfect fluids that can be obtained out of the general EMT \eq{genEMT} can be classified according to their conserved currents, see also \cite{Ballesteros:2016gwc}. These currents are of two types: Noether currents and currents that are conserved independently of any symmetry of the action. In what follows when we say that a current is conserved we mean that it is {\it covariantly} conserved, i.e.\ $\nabla_\mu J^\mu=0$.
\begin{enumerate}
\item[\textbf{\textit{i)}}] $U(b)$ has currents of two types. First, there is 
\begin{align} \label{ec}
\mathcal{J}^\mu =b\, u^\mu\,,
\end{align} 
which is conserved off-shell. Actually, any current of the form $f(\Phi^a)\mathcal{J}^\mu$, where $f$ is a function of the fields $\Phi^a$, is also conserved thanks to $u^\mu\partial_\mu\Phi^a=0$. This last equation shows that $\Phi^a$ are actual comoving coordinates of the fluid, as the pull-back formalism requires. 
\item[\textbf{\textit{ii)}}] $U(X)$ has only one independent conserved current:
\begin{align} \label{xmu}
\mathcal{X}^\mu=-2\sqrt{-X}\,U_X\,\mathcal V^\mu\,,
\end{align}
coming from the symmetry $\Phi^0\rightarrow \Phi^0+c^0$, $\partial_\mu c^0=0$, and giving the dynamics of $\Phi^0$ for $U(X)$. The factor $2$ is included for later convenience.
\item[\textbf{\textit{iii)}}] $U(Y)$ has, in addition to the same currents as $U(b)$, the current 
\begin{align} \label{ymu}
\mathcal{Y}^\mu=U_Y\, u^\mu\,.
\end{align}
As a matter of fact, any $f(\Phi^a)\, \mathcal{Y}^\mu$ is conserved as well. These currents are due to the symmetry $T_s$: $\Phi^0\rightarrow \Phi^0+f(\Phi^a)$. 

\item[\textbf{\textit{iv)}}] $U(b,Y)$ has the same currents as $U(Y)$ because they share field content and symmetries. We recall that $U(b,Y)$ is the most general case at LO assuming that the action is symmetric under $V_s$Diff and $T_s$. 
\end{enumerate}
As we will see in Section, \ref{interp} only the currents
$\mathcal{J}^\mu$, $\mathcal{Y}^\mu$ and $\mathcal{X}^\mu$ are needed
to establish the full thermodynamic dictionary.
Actually, other conserved currents are present (depending on the operator content)  but are not relevant for
our purposes in this work; for a discussion see for instance~\cite{Dubovsky:2005xd,Endlich:2010hf, Dubovsky:2011sj,Ballesteros:2012kv,Ballesteros:2016gwc}.
Requiring only the symmetry $V_s$Diff and the shift symmetry \eq{shift} for $\Phi^0$  the action is \eq{genfluid} and the EMT is not a perfect fluid. In this case, the conserved current associated to the second of these symmetries is precisely the sum of the currents \eq{xmu} and \eq{ymu}: $\mathcal{X}^\mu+\mathcal{Y}^\mu$. As before, any current of the form $(\mathcal{X}^\mu+\mathcal{Y}^\mu)\,f(\Phi^a)$ is also conserved, by virtue of $u^\mu\partial_\mu\Phi^a=0$.

\subsection {Energy density and pressure of perfect fluids}
\label{sect:baro}
Barotropic fluids are common in several branches of physics; in particular in cosmology. They serve to model in a first and crude approximation the basic matter species of the $\Lambda$CDM model: cold dark matter (CDM), baryons, photons and neutrinos. The current accelerated expansion can also be modeled with a barotropic fluid, be it a cosmological constant, $\Lambda$, or a more exotic component with a sufficiently negative {\it equation of state} $w=p/\rho$. The usual definition is that barotropic fluids are those whose pressure is a function of the energy density alone, i.e.\ $p=p(\rho)$. This can be generalized by defining a barotropic fluid as one whose pressure is completely characterized knowing the energy density at each point of the fluid (or vice versa) \cite{Ballesteros:2012kv}. In this sense, the perfect fluids $U(X)$, $U(b)$ and $U(Y)$ are all barotropic. Indeed, in these cases the pressure and the energy density are related, respectively, through the relations: 
\begin{align} \label{LT}
\begin{aligned}
i) &\quad \rho=-U(b)\,,  \quad \,& \quad p=U-b\,U_b\,,\\
ii) &  \quad p=U(X)\,,\quad  & \rho=2\,X\,U_X-U\,,\\
iii) & \quad p=U(Y)\,,\quad  & \rho=Y\,U_Y-U\,. 
\end{aligned}
\end{align}
The  perfect fluids $U(b,Y)$ are not barotropic in general, simply because their Lagrangians are functionals of two independent operators:
\begin{align} \label{UbY}
iv)\quad p=U-b\,U_b \quad \text{and} \quad \rho=Y\,U_Y-U \,, \quad U=U(b,Y)\,.
\end{align}
The relation between pressure and energy density for $U=U(b,Y)$ can be
determined only if another thermodynamic quantity is also
known. However, there exist specific choices of the function $U(b,Y)$
that are barotropic. In particular, a constant $w=p/\rho$ can be
obtained for $U(b,\,Y)=b^{1+w} \;{\cal U}(Y\,b^{-w})$, with $\cal U$ being
an arbitrary function. Another possibility is  $U(b,Y)=b^{1+w}+\lambda
\;Y^{{1+1/w}}$, 
with constant $\lambda$. A constant equation of state can also be reproduced with $U(X)=X^{{(1+1/w)/2}}$. Nonrelativistic matter with $p=0$, such as standard CDM, admits only the description $U(b)=b$.

There are also other examples of perfect barotropic fluids that have been used often  in astrophysics and cosmology, and can be easily described within our framework. For instance, a Chaplygin gas satisfies  $p=-{A}/{\rho}$, with constant $A$ and can be obtained from $U(b)=\sqrt{A+\lambda\, b^2}$, $U(Y)=\sqrt{A+\lambda\;Y^2}$ or $U(X)=\sqrt{A+\lambda\;X}$, with constant $\lambda$.

\section{Thermodynamic dictionary for perfect fluids} \label{interp}
\label{sect:dict}
In this section  we give the thermodynamic interpretation of the EFT
of perfect fluids described in Section \ref{sect:fluida} by giving the correspondence between the thermodynamic variables and the
operators $b$, $Y$ and $X$ for the different kinds of perfect fluids. These results extend the interpretations that were previously proposed in \cite{Dubovsky:2005xd,Dubovsky:2011sj}.

For a simple thermodynamic system, among the six variables 
($s,\,T,\,n,\,\mu,\,\rho,\,p$), two of them  can be taken as
independent, say $z_1$ and  $z_2$, and  the remaining four, can be expressed in terms
$z_1$ and $z_2$. The EMT provides   the explicit form of $p$
and $\rho$; see (\ref{LT}-\ref{UbY}).
Our outcome, once the independent variables have been chosen, is that
$U$ is proportional to the natural thermodynamical potential
associated with such thermodynamical variables; see  Table \ref{tab:Leg}.
In Appendix~\ref{app} we describe a method based on the counting of
derivatives of $U$ with respect  to the EFT operators, showing that this is indeed the case.

Let us start with the fluid
described by  $U(b,Y)$, which depends on two operators, and choose as
independent thermodynamic  variables $(T, \, n)$. From (\ref{UbY}) we have that  
$\rho = Y \, U_Y - U$ and  $p=  U - b \, U_b$.  
If a thermodynamic
interpretation exists, the  operators $b$
and $Y$  of the scalar field theory should be functions of the
chosen  thermodynamic  variables, namely  $Y=Y(T,n)$ and $b=b(T,n)$.
From Table \ref{tab:Leg} it is clear that the natural thermodynamic
potential with   variables $(T, \, n)$  is the free-energy density ${\cal F}$,
which is obtained from $\rho$ or $p$ after a single Legendre
transformation. The natural choice is to take
$U(b(n,T),Y(n,T))=-\mathcal{F}(n,T)$ and find for which  $b=b(n,T)$
and $Y=Y(n,T)$  the fundamental thermodynamic relations \eq{termo1}
and \eq{termo2} are satisfied.
There are multiple ways in which this can be done, all of them leading
to the same set of solutions. Using that $d\mathcal{F} = \mu\, dn -
s\, dT$ -- see Table~\ref{tab:Leg}--, the Euler relation \eq{termo2}
becomes
$b\,\mathcal{F}_b-Y\mathcal{F}_Y=n\,\mathcal{F}_n-T\,\mathcal{F}_T$. Expressing
the derivatives with respect to $b$ and $n$ in terms of derivatives
with respect to $T$ and $n$ and imposing that the resulting equation
has to be valid for all $\mathcal{F}$, we obtain $b=n\,b_n-T\,b_T$ and
$Y= T\,Y_T-n\,Y_n$. We can now use the Gibbs-Duhem relation
\eq{termo3} or, simply, the definition $\mathcal{F}=\rho-T\,s$. Either
way, we are led to $T\propto Y$, $n\propto b$, $s\propto U_Y$ and
$\mu\propto U_b$; see Table \ref{tab:thermo}. An analogous procedure
can be followed when the thermodynamical variables $(s, \, \mu)$ are
used, in this case the natural thermodynamic potential  is ${\cal I}$ and one
can check that setting $U(b(s,\mu),Y(s,\mu))=-\mathcal{I}(s,\mu)$ all the
thermodynamic relations are satisfied if $b=s$ and $Y=T$; the result is also given in Table \ref{tab:thermo}. 
\no
\begin{table}[t!]
  \footnotesize
%  \tiny
  \renewcommand\arraystretch{1.35}
  \centering  
  \begin{tabular}{||c||c|c|c|c||}
    \cmidrule{2-5}\morecmidrules\cmidrule{2-5}
    \multicolumn{1}{c||}{}
 %Potentials    
& \cellcolor{gray!8} $\boldsymbol{U(b)}$& \cellcolor{gray!8} $\boldsymbol{U(Y)}$ & \cellcolor{gray!8} $\boldsymbol{U(X)}$ & \cellcolor{gray!8} $\boldsymbol{U(b,\,Y)}$    \\ 
  \multicolumn{1}{c||}{}    & $\rho=-U$ &  
             $\rho=-U+ Y\,U_Y$ &
            $\rho=-U+2\,X\,U_X $ & 
            $\rho=-U+ Y\,U_Y$ \\
            % Independent 
      \multicolumn{1}{c||}{}       & $p=U-b\;U_b$  & $p=U$ &$p=U$ &$p=U-b\;U_b$
               \\ 
           % Variables      
  \multicolumn{1}{c||}{}          & $\mathcal{J}_{\mu}=b\;u_{\mu}$  & ${\cal Y}_{\mu}=U_{Y}
                \;u_{\mu}$ &${\cal X}_{\mu}=-2\,(-X)^{1/2}\,U_X\;{\cal
                             V}_{\mu}$ &$\mathcal{J}_{\mu}=b\;u_{\mu},\;{\cal Y}_{\mu}=U_{Y}\;u_{\mu}$
               \\ \hline  \hline  
    \multirow{4}{*}{\begin{tabular}{@{}c@{}}$ \boldsymbol{(\mu,\,s)}$ \\ ${\cal I}=-U$\end{tabular}}  & $b=s$ & 
                           $Y=\mu$ &
                           $X=-\mu^2$& 
                           $b=s,\;Y =\mu$ \\
                     &   $n=0 $&
                       $n=U_Y  $&
                       $n=-2\,U_X\,\sqrt{-X} $&
                       $ n=U_Y   $\\
                    &   $ T=-U_b$&
                       $ T=0$&
                       $ T=0 \, , \quad {\cal X}_\mu =-  \, n/2\, {\cal V}_\mu$&
                       $ T=-U_b  $\\ 
              &  $\mathcal{J}_{\mu}=s\;u_\mu $  &
                 ${\cal Y}_\mu=n\;u_\mu $&
                       $  {\cal X}_\mu =\, n\, {\cal V}_\mu$&
                       $ \mathcal{J}_{\mu}=s \, u_{\mu} \, , \; {\cal Y}_{\mu}=n \, u_{\mu}  $\\
                        \hline  \hline  
\multirow{4}{*}{\begin{tabular}{@{}c@{}}$ \boldsymbol{(n,\,T)}$  \\ ${\cal F}=-U$\end{tabular}}  & $b=n  $ & 
                           $Y=T$ &
                           $X=-T^2$& 
                           $b=n,\;Y =T$ \\
                    &   $s=0 $&
                       $s=U_Y$&
                       $s=-2\,U_X\,\sqrt{-X} $&
                       $ s=U_Y    $\\
                    &   $ \mu=-U_b$&
                       $ \mu=0$&
                       $ \mu=0 \,  $&
                       $  \mu=-U_b    $\\
                    &   $ \mathcal{J}_{\mu}=n\,u_\mu$&
                      $ {\cal Y}_\mu=s\,u_\mu$&
                       $  \quad {\cal X}_{\mu}=s \, {\cal V}_{\mu}$&
                       $ \mathcal{J}_{\mu}=n\,u_{\mu} \, , \; {\cal Y}_{\mu}=s\,u_{\mu} $\\  \hline
   \hline
   \multirow{4}{*}{\begin{tabular}{@{}c@{}}$ \boldsymbol{(\mu,\,T)}$   \\ $\omega =-U$\\ \\$z={\mu/T}$ \end{tabular}}  & \cellcolor{gray!5}
&  $Y=T \, f \left(z\right)$ &  $X=-T^2\; f\left(z\right)$& \cellcolor{gray!5}  \\
                    &  \cellcolor{gray!5}  &  $s=U_Y\,(f - z \, f')$&$s=U_X\,(\mu\,f'-2\,T\,f) $&\cellcolor{gray!5} \\ 
                     &  \cellcolor{gray!5} &  $  n=f' \, U_Y$&$n=-U_X\,T\,f' $&\cellcolor{gray!5}\\ 
                     &  \cellcolor{gray!5} & $T f  {\cal Y}_{\mu}=
                             (\rho+p)\,u_\mu $&$T \,f^{1/2}
                                                                      {\cal
                                                                      X}_{\mu}=(\rho+p){\cal V}_{\mu} $& \cellcolor{gray!5} \\  \hline \hline
      \multirow{4}{*}{\begin{tabular}{@{}c@{}}$  \boldsymbol{(n,\,s)}$  \\ $\rho=-U$\\ \\$\sigma=s/n$ \end{tabular}}   
     &  $b=s\, f\left(\sigma^{-1}\right)$                     &\multicolumn{ 3}{c||}{\cellcolor{gray!5}}   \\
                     &  $\mu=-U_b\,f' $   &\multicolumn{ 3}{c||}{\cellcolor{gray!5}} \\
                    &  $ T=-U_b\, (f- f'/\sigma)$ &\multicolumn{ 3}{c||}{\cellcolor{gray!5}} \\
                    &  $
                 \mathcal{J}_{\mu}=s\, f\,u_\mu$   &\multicolumn{ 3}{c||}{\cellcolor{gray!5}}  \\
 \hline  \hline 

  \end{tabular}
  \caption{ \small  \it The EFT-thermodynamics dictionary for perfect
    fluids. The first row gives the energy density, the pressure and
    the relevant currents for the thermodynamic interpretation of each
    type of perfect fluid. The other rows give the entries of the
    dictionary for the four pairs of independent thermodynamic
    variables and their corresponding potentials. When it appears, $f$
    is an undetermined function (and $f'$ is its derivative) of the argument
   % We have  also defined 
   $z=\mu/T$ or $\sigma=s/n$} 
  \label{tab:thermo}
\end{table}
The possibility of identifying $-U(b,Y)$ as the thermodynamic
potential $\mathcal{I}(s,\mu)$ was already found in
\cite{Dubovsky:2011sj}.

The dictionary of Table \ref{tab:thermo} shows that some combinations of operator content and independent thermodynamic variables have no entry.
By looking at Table \ref{tab:thermo}, for the case $U(b, \, Y)$, note that in the dictionary there is no (direct)
entry for some choice of independent thermodynamical variables. 
This
is the case for the choice of variables $(\mu, \, T)$ and  $(n, \,
s)$ for $U(b,Y)$. The reason is that the EMT tensor for $U(b, \, Y)$ is such that
$U$ is not proportional to $p= -\omega$ or to $\rho$ as
thermodynamics should require (see Table \ref{tab:Leg}) if $U$ is
interpreted to be the corresponding thermodynamical potential. 
As it was already argued in \cite{Dubovsky:2011sj}, $U(b,Y)$ can be identified with $-I(s,\mu)$ because this potential can be obtained from $\rho(n,s)$ using a Legendre transformation; see Table \ref{tab:Leg}. The identification of $U(b,Y)$ with $-\mathcal{F}(n,T)$ can be argued on the same ground.

It is important to stress that the nonexistence of an entry in the
dictionary for some variables  does not mean that it is impossible to use those variable. Take, for instance, the
entries relative to $U(b , \, Y)$ and variables $(n, \, T)$. By using a
Legendre transformation, one can  always switch from  $(n, \, T)$ to  $(n, \,
s)$, still getting $T=Y$ and $\mu =- U_b$. Thus the entries in the
dictionary correspond to  {\it inequivalent} 
thermodynamic interpretations of the very same 
scalar field action. 

Let us now consider the Lagrangian $U(b)$. In this case given the fact
that at least two independent thermodynamical variables are needed to describe a simple system and a
single operator is present in the action, it is not
surprising that various possibilities exist. Using the Euler
relation \eq{termo2} and the expressions \eq{LT} for the energy
density and pressure, we get $\mu\,n+T\,s=-b\,U_b$. In addition, the
Gibbs-Duhem equation \eq{termo3} tells us that
$s\,dT+n\,d\mu=-b\,U_{bb}\,db$, 
and  the first principle of thermodynamics \eq{termo1} is
$T\,ds+\mu\,dn=-U_b\,db$. To proceed further we can choose $n$ and $T$
as our two independent thermodynamic variables, so that $b=b(n,T)$. It
is easy to check that $n=b$ and and $\mu=-U_b$ solve these equations
for all $U_b$. 
This implies that $s=0$ and, according to the fourth postulate of Section \ref{sect:basics}, $T=0$. 
Therefore $U(b)$ can describe a perfect fluid in the limit of zero temperature. 
With this interpretation, that was already proposed in \cite{Dubovsky:2005xd}, the current $\mathcal{J}^\mu$ defined in \eq{ec} plays the role of the conserved particle current, and the entropy current vanishes in this limit. 

Let us consider now as independent thermodynamic variables
$(\mu,\,  T)$, i.e. we assume $b=b(\mu,\,T)$. It is easy to see that $b=s$
and $T=-U_b$ can solve the thermodynamic equations. For consistency,
this implies that $n$ must be zero. According to the discussion in
Section \ref{sect:basics}, the choice $n=0$ should be understood as
zero particle density. However, in relativistic hydrodynamics $n$ is
usually meant to represent a charge density, making this point of view
more appealing. In this case, $b=s$ and the current  $\mathcal{J}^\mu$
represents the entropy current. This is the interpretation that was earlier
advocated in \cite{Dubovsky:2011sj} and subsequent works. 

It is interesting to note that there are choices of independent
thermodynamic variables for which an arbitrary function of their ratio
appears in the dictionaries of $U(b)$.  Consider as an example $U(b)$
and take $n$ and $s$ as independent variables, so that the natural
thermodynamic  potential is the energy density, $\rho(n,s)$. The equation $\rho+U=0$
is automatically satisfied, whereas the  Euler  relation becomes 
$b = n \, b_n + s \, b_s$, whose  general solution is $b = s \, f
({n}/{s})$, with $f$ being  an unspecified function of $\sigma^{-1}=n/s$. The
conservation of the  current $\mathcal{J}^\mu = b\, u^\mu$ gives
$(\rho+p)\theta=-T\,s' -s\,T'$, which is the first principle of
thermodynamics in the form 
\eq{ns}. Similarly, for $U(X)$ and $U(Y)$ with $\mu$ and $T$ as
independent thermodynamic variables, the corresponding entries of the
dictionary depend on the ratio $z=\mu/T$.  An interesting discussion of the
case $U(X)$ can be found in~\cite{Matarrese:1984zw}.  When an
unspecified function $f$ is found, it also enters in the conserved
currents and their physical interpretation is more subtle.

Notice  that the entries for $U(b)$ and $U(Y)$ relative to the
variables $(\mu, \, s)$ and $(n, \, T)$ represent also the limiting
cases of $U(b, \, Y)$  for $U_Y=0$ and $U_b=0$ respectively. 
In addition, the limits for $(\mu,\,s)$ and $(n,\,T)$ can be obtained choosing  specific functions $f$ 
for the fluids that depend on a single operator. Concretely, this can be done choosing $f$ to be a constant or, imposing $f=n/s$ for $U(b)$; $f=\mu/T$ for $U(Y)$ and $f=\mu^2/T^2$ for $U(X)$.

The analysis of this section leads to the conclusion that $U(b , \,
Y)$ arises as the most appropriate Lagrangian for a complete
thermodynamic description of a relativistic perfect fluid. Having two
effective operators  allows a full matching to the thermodynamic
relations describing a  simple system.  This is translated into the
fact that for $U(b,Y)$, the currents $\mathcal{J}^\mu$ and
$\mathcal{Y}^\mu$ can be put in correspondence with the entropy and
particle currents --both of which are conserved-- giving an adiabatic
fluid, see Section \ref{sect:basics}. The Lagrangian $U(b,Y)$ was
already identified in \cite{Dubovsky:2011sj} as the most general one
for a perfect fluid carrying a conserved charge. Moreover, it is
remarkable that $U(b,Y)$ is selected from the EFT of continuous media
at LO in derivatives by the symmetry $V_s$Diff\,$\times\, T_s$; see
Figure \ref{fig:cartoon}. In addition, thanks to this symmetry,
$U(b,Y)$ includes all four St\"uckelberg fields needed for a full
embedding of the fluid in spacetime, but without introducing two
different four-vectors; see Section \ref{sect:fluida} and also
\cite{Ballesteros:2016gwc}. It is also worth pointing out that
$U(b,Y)$ is a nonbarotropic (and nonisentropic) fluid, which opens
the possibility of 
describing the dynamics and thermodynamics of a broad variety of
physical systems. In summary, for a complete thermodynamic 
description of perfect fluids at low energies as simple systems, the
standard pull-back formalism must be extended with an extra 
scalar, $\Phi^0$, and with  the symmetry $V_s$Diff\,$\times\, T_s$.

\section{Thermodynamics with broken shift symmetry?}

In this section we argue that the shift symmetries \eq{shift} are essential for a consistent thermodynamic interpretation of the EFT of perfect fluids. Generically, it is sufficient that the action of the four St\"uckelberg fields does not respect one of these symmetries to prevent a proper thermodynamic description.
To illustrate this point we will consider the simplest possible example, assuming that the shift symmetry \eq{shift} of the case
$U(X)$ is broken by the explicit appearance of $\Phi^0$ on the master
function. So, we consider a {\it k-essence} Lagrangian
\cite{ArmendarizPicon:1999rj,ArmendarizPicon:2000dh,Deffayet:2010qz,Pujolas:2011he}, given by
$U(X,\,\Phi^0)$. The (nonvanishing) covariant divergence of the
current ${\cal X}^\mu=-2\,\sqrt{-X}\,U_X\,{\cal V}^\mu$ is the
equation of motion for $\Phi^0$, i.e.\ $\nabla^\mu  {\cal X}_\mu=U'$, where $U'=\partial_{\Phi^0}U$.
The EMT is formally of the same form as for $U(X)$, thus describing a perfect fluid with $\rho = 2 \, X \, U_X - U$, $p = U$ and four-velocity $\mathcal{V}_{\mu}$. From the relations \eq{T4} and (\ref{ns}) we find
\be
s_\nu\,\nabla^\nu T\,+ n_\nu\,\nabla^\nu \mu =
\sqrt{-X}\,U'+{\cal X}^\mu\,\nabla_\mu\sqrt{-X} \, .
\label{up}
\ee
Let us now attempt to construct a thermodynamic interpretation of {\it
  k-essence} by considering the various possible combinations of two
 independent variables among $n$, $T$, $\mu$ and $s$, in the same way as in the previous section. 
\begin{itemize}
\item $(\mu,\,s)$: In this case we get $X=-\mu^2$ and $\Phi^0$
is independent from $\mu$ and $s$. We find that $T=0$ and ${\cal X}^\mu=n\,u^\mu$, so that (\ref{up}) implies
that  $U'=0$. 
\item  $(n,\,T)$: Then $X=-T^2$ and $\Phi^0$ is
independent from $n$ and $T$. Besides,  $\mu=0$ and ${\cal X}^\mu=s\,u^\mu$. Again, (\ref{up}) implies that $U'=0$.  
\item $(\mu,\,T)$: 
Taking $U(X(\mu,T)=p(\mu,T)$,  from the Euler and the Gibbs-Duhem relations we obtain that $X=-\mu^2\;f_1({T}/{\mu})$ and 
$\phi=f_2({T}/{\mu})$ with $s=-f_1'\,U_X\sqrt{-X/f_1}\,+f_2'\,U'/\sqrt{-X/f_1}$ and $n=(f_1'\,T-2\,\sqrt{-X\,f_1})\,U_X+f_2'\,f_1\,T\,U'/X$. Since the correspondence must be valid for all $U(X,\Phi^0)$, the equation \eq{up} implies that $U'=0$. 
\end{itemize}
The thermodynamic relations are incompatible with the equations of
motion unless the Lagrangian does not depend on $\Phi^0$. We conclude
that {\it k-essence} only 
admits a thermodynamic interpretation if the shift symmetry is enforced, that is: if the action depends only on $X$. 
The above statement can be easily extended to the other cases.

 \section{Some simple applications of the thermodynamic dictionary}\label{sect:app}
 
 \subsection{Bose-Einstein and Fermi-Dirac distributions}
Consider the Bose-Einstein (BE) and Fermi-Dirac
(FD) statistics for (noninteracting) particles of mass $m$ with spin degeneracy $g$ per energy state. The number of particles per unit of momentum $k$ at temperature $T$ is a function of the  chemical potential $\mu$:
\be
2 \pi^2\,F(k,\,T,\mu)=\frac{g}{\exp\left[(\sqrt{k^2+m^2}-\mu)/T\right]
  +\epsilon}\,,\qquad\epsilon=\pm1
\label{qfg}
\ee
In this expression $\epsilon =1$ corresponds to FD and $\epsilon =-1$
to BE. For convenience we have set the Boltzmann constant to be 1.
Some interesting limits are controlled by the ratios $m/T$ and
$z=\mu/T$. If $z\gg 1$ the gas becomes degenerate; whereas if $z\ll 1$
the particles have more freedom to occupy higher energy levels. The
deeply relativistic limit  is $T\gg m$  and, conversely, the gas
becomes nonrelativistic at sufficiently low temperatures, i.e.\ $m\gg T$. 

For $T\gg m$ we get
\be
\label{ultrel}
p=\frac{\rho}{3}=- \epsilon \,\frac{g}{ \pi^2}T^4 \, \text{Li}_4\left(-\epsilon\, \alpha
   \right) \, , \qquad n=- \,\epsilon\,
 \frac{g}{ \pi^2}\,T^3\,\text{Li}_3\left(-\epsilon\, \alpha \right) \, , \qquad
 s=\frac{4}{3}\frac{\rho}{T}-n \, \log(\alpha)  \, ;
\ee
where $\text{Li}_x(z)$ is the polylogarithm function of order $x$ \cite{askey1982} and $\alpha =\exp(z)$ is the {\it fugacity}.
Using Table~\ref{tab:thermo}, we can reproduce these relations by
choosing $(\mu, \,T)$ and taking
\be
U(Y)=\frac{g}{3} \, Y^4,\quad Y=T\, f(z)=T\left[-\frac{\epsilon}{\pi^2}\text{Li}_4\left(-\epsilon \,
   e^z \right)\right]^{1/4} \; .
\ee
The high temperature limit of a deeply relativistic gas is obtained taking
$\alpha\to 0$, which gives $Y\propto  \, T$. In the low temperature limit, $\alpha \to \infty$, the gas is degenerate and
$\text{Li}_n\left(-\epsilon \, \alpha \right) \to
\frac{\alpha^n}{n\,\epsilon}$.  As a result,  $Y\propto\mu$. The same conclusions can be reached by using instead
$U={g}\,X^2/3$, with $X=T^2\,f_2(\alpha)$ and
$f_2(\alpha)=\left[\frac{-\epsilon}{\pi^2} \,\text{Li}_4\left(-\epsilon \, \alpha
   \right)\right]^{1/2}$.
\vskip .5cm
\no

\no
In the nonrelativistic limit $m \gg T$, the expressions (\ref{qfg}) become
\ba n=g \, \left(\frac{m \, T}{2 \pi}\right)^{3/2}\, \exp\left({\frac{\mu-m}{T}}\right)\,, \qquad 
 \rho=n \, \left(m+\frac{3 \, T}{2}\right),\qquad p=n \, T\ll\rho \, . \label{rholim} 
\ea
These  can be reproduced with $U(b, \, Y)$, being $b=n$, $Y=T$,  and taking 
\be
U(b, \, Y)= b \, Y\left\{ 1+ \text{Log}\left[\frac{g}{b} \, \left(\frac{m
        Y}{2 \pi}\right)^{3/2}  \right] \right\} - b \, m\, . 
%b  - p(Y)\quad \text{with }  Y=T \, ;
\label{ideal}
\ee

When both $\alpha $ and $m/T$ are non-negligible, getting a relation among the EFT operators $b$ and $Y$ and the thermodynamic variables involves 
solving an integrodifferential equation for $U(b,Y)$, using $b = n(T, \mu)$ and $\mu  =-U_b+Y\,U_{bY}$.
\subsection{van der Waals gas}
In this case the pressure and the particle number density are related by: $(p+\kappa \,n^2)\,(1-\gamma\,n)= \,T\,n$, 
where the constants $\kappa$ and $\gamma$ model the finite
molecular volume of the gas and small intermolecular interactions, respectively.  This equation of state can be easily obtained from $U(b,Y)$ choosing the pair $\{n,\,T\}$ as independent thermodynamic variables, so $b=n$ and $Y=T$. Integrating $p=U-b\, U_b$ we obtain $U=b \left[ b \, \kappa +Y \log
   \left(b^{-1}-\gamma\right)+{\cal U}(Y) \right]$. The function ${\cal U}(Y)$ can be determined imposing that for $\kappa=0$ and $\gamma=0$ we recover the expression (\ref{ideal}) for
the ideal gas, which leads to
\be
U(b, \, Y) = b \left \{ b \, \kappa + Y - m + Y  \log \left[
    \frac{g \left(1-\gamma\,b\right)}{b} \left( \frac{ m \, Y}{ 2 \,
        \pi}\right)^{3/2} \right] \right \} \,,
\ee
and hence
\begin{align}
\begin{aligned}
\rho &=  b \left( m+ \frac{3}{2} \, T - n \, \kappa \right) \,,\\
s & = \frac{5 \, b}{2} + b \,\log \left[ \left( \frac{ m \, Y}{ 2 \,
        \pi}\right)^{3/2} \, \frac{g(1- b \, \gamma)}{b}
\right] \,,\\
\mu & = m- 2 \, b \, \kappa + \frac{b \, Y \gamma}{1- b \, \gamma} + Y
\, \log \left[ \frac{b}{g(1- b \, \gamma)} \left( \frac{ m \, Y}{ 2
      \,\pi}\right)^{-3/2}  \right] .
      \end{aligned}
      \end{align}

\subsection{Polytropic fluids}

Polytropic fluids are used to model the behavior of 
matter under a wide range of physical conditions, including e.g.\
the interior of neutron stars. It is convenient  to separate  the  mass contribution $\rho_0= n \, m$
($m$ is the individual mass of the fluid's constituents) and the
internal energy density $\epsilon_I$ from the energy density: 
$\rho=\rho_0+\epsilon_I$. The equation of state then reads
$p=\kappa\;\rho_0^{\Gamma}$, with $\Gamma$ constant. 
A polytropic equation of state can be described by
$U(b)=\lambda \, b+\kappa \, \frac{b^{\Gamma}}{1-\Gamma}$,
taking $\{n , \, T\}$  as thermodynamic variables. In this case $b=n$ and $\rho_0=b \, m$, $p(b)=U-b\, U_b=\kappa\,b^{\Gamma}$.
Polytropic equations of state can also be described by $U(Y)$ and
$U(X)$, though  the expressions are more involved and will be omitted.

\subsection{Ehrenfest-Tolman effect}
If the spacetime curvature is nonzero and there exists a timelike Killing vector $\xi^\mu$, the equilibrium
temperature  $T$ satisfies --according to the Ehrenfest-Tolman effect-- the relation 
\be  \label{cond}
T \, \sqrt{-g_{\mu \nu} \xi^\mu \xi^\nu} = \text{constant} \, .
\ee
An analogous relation holds as well  for the chemical potential $\mu$,  see for instance \cite{LandauStat}.
The temperature $T$ and the chemical potential $\mu$ of our
dictionary --see Table \ref{tab:thermo}-- are consistent with the Ehrenfest-Tolman result provided that $v^\mu \propto \xi^\mu$ and $\xi^\mu \partial_\mu
\Phi^0=1$. Indeed, $T$ and $\mu$ can be
associated to $Y$ and $-X^2$, respectively. For a static spacetime, the
timelike Killing  vector $\xi^\mu$  can be identified with $\partial^\mu
\Phi^0$ and thus (\ref{cond}) holds, showing that the Ehrenfest-Tolman
effect takes place.

\section{Thermodynamic stability}
\label{sect:stab}
Thermodynamic stability of a system requires that the Hessian matrix of the function $s(\rho,\,n)$ must  be negative definite,\footnote{A 
 $2\times 2$ matrix $M$  is negative definite if $\text{Tr}(M)\leq 0$ and $\text{det}(M)\geq 0$. Conversely, it
 is positive definite if $\text{Tr}(M)\geq 0$ and $\text{det}(M)\geq 0$.} which is guaranteed by the conditions
\be
\label{stabs}
s_{\rho\rho}+ s_{nn}\leq0 \quad {\rm and}\quad  s_{\rho\rho}\,s_{nn}-s_{\rho n}^2\geq 0\,.
\ee 
These equations can also be equivalently formulated in terms
of conditions for the
 energy density, whose Hessian must be positive
definite: 
\be
\rho_{ss}+\rho_{nn}\geq 0\quad {\rm and}\quad
\rho_{ss}\,\rho_{nn}-\rho_{sn}^2\geq 0 \, .
\ee

It is also possible to use the other potentials of Table \ref{tab:Leg}. In particular, for the free-energy density ${\cal F}(n \, , T)  =\rho-T\,s$, the conditions reduce simply to ${\cal F}_{nn}\geq 0$ and ${\cal F}_{TT}\leq 0$ (while the condition on the mixed second derivative is automatically implied by these two). Similarly, for the potential $\mathcal{I}(\mu,s)=\rho-\mu\,n$ the conditions are ${\cal I}_{\mu\mu}\leq 0$ and ${\cal I}_{ss}\geq 0$. If any of these potentials ($\mathcal{F}$ or $\mathcal{I}$) are identified with $-U$, which according to Table \ref{tab:thermo} occurs only for $U(b,Y)$, the thermodynamic stability conditions are translated into simple constraints on the derivatives of $U$. Concretely:
\begin{align} \label{ts1}
U_{bb}\leq 0 \quad \text{and}\quad  U_{YY}\geq 0\,.
\end{align}

Similarly, for the grand potential $\omega(\mu \, , T)=-p= \rho - s \,
T - n \,\mu$, thermodynamic stability is guaranteed when
\be
\omega_{\mu\mu}+ \omega_{TT} \leq 0 \, , \qquad
\omega_{\mu\mu}\,\omega_{TT}-\omega_{\mu T}^2\geq 0 \, .
\ee
If $-U$ is identified with  $\rho$ or $\omega$, as it is the case for the barotropic fluids that depend only on $X$, $b$ or $Y$,  the thermodynamic dictionary involves an undetermined function, $f$, of a ratio of independent thermodynamic variables; see Table \ref{tab:thermo}. In these cases, thermodynamic stability leads not only to conditions on the derivatives of $U$,  but also to some constraints involving derivatives of $f$. Concretely, for $\rho=-U(b)$ we get 
\be
%$
U_{bb}\leq 0%$
\;\;\; {\rm and}\;\; \;% $
f''\,U_b\leq 0 %$
,
\ee
 where primes denote derivatives of $f$ with respect to its argument; see Table \ref{tab:thermo}. Similarly, for $\omega = -U(Y)$ we obtain 
 \be%$
 \label{uf}
 U_{YY}\geq 0%$
 \;\;\;{\rm and}\;\;\; %$
 f''\,U_Y\geq 0 .%$
 \ee Finally, in the case $\omega =-U(X)$, we get: 

\begin{align} \label{ts2}
2\,X\,
  U_{XX}+U_X \leq0\,,\;\;\;{\rm and}\;\;\;
((f')^2-2\, f\,  f'')\, U_X\geq 0.
  \end{align} 

If we take the case of the free  relativistic BE or FD  distribution (see Section \ref{sect:app}) we can specify the function $f(z)$,
where $z=\mu/T$, and then we get:
 $f''(z)\geq0$  when $f\sim(-\epsilon \,{\rm Li}_4(-\epsilon\,e^z))^{1/4}$ (which is the case for the $U(Y)$ and it implies $U_Y\geq0$) while  $((f')^2-2\, f\,  f'')\leq0$ when
$f\sim(-\epsilon \,{\rm Li}_4(-\epsilon\,e^z))^{1/2}$ (for the $U(X)$ case and it implies $U_X\leq0$). 
\footnote{In  the BE case ($\epsilon=1$)  the chemical potential $\mu$ is negative  and $0\leq \alpha\leq 1$.}

\section{Sound speed} \label{sound}

We can define the sound speed of perfect fluids as the quantity that relates the variations of the energy density and the pressure along the fluid flow:
\begin{align} \label{so}
p'=c_s^2\,\rho'.
\end{align} 
As we will see in the next section, where we discuss dynamical stability, this definition gives the speed of propagation of longitudinal phonons, appearing naturally by expanding the action at quadratic order in fluctuations. For barotropic perfect fluids, \eq{so} can be simply computed as $c_s^2=dp/d\rho$ using the expressions \eq{LT},  giving
\be
c_s^2 = \begin{cases}
b\;\frac{U_{bb}}{U_b}\,, & U=U(b) \\[.2cm]
\frac{U_{Y}}{Y\;U_{YY}}\,, &  U=U(Y)\,\quad.\\[.2cm]
\frac{U_{X}}{U_X+2\,X\,U_{XX}}\,, & U=U(X)
\end{cases} % \;\;
\label{speeds}
\ee

In the case of $U(b,Y)$,  a variation of the pressure is not uniquely determined by the variation of the energy density. This is simply because $U(b,Y)$ is a nonbarotropic fluid, 
 see \eq{UbY}. 
In this case, it is convenient to define the restricted variations of the pressure with respect to the energy density
 \be
\label{ci}
 c_b^2=\left. \frac{\partial p}{\partial \rho}\right |_b =\frac{U_Y-b\,U_{bY}}{Y\,U_{YY}}\, ,\quad
 c_Y^2=\left. \frac{\partial p}{\partial \rho}\right |_Y =\frac{b\,U_{bb}}{U_b-Y\,U_{bY}}\,.
 \ee  
To obtain the relation between the variations of the energy density and the pressure we make use of  the conservation of the currents $\mathcal{J}^\mu=b\,u^\mu$ and $\mathcal{Y}^\mu=U_Y\, u^\mu$, defined in \eq{ec} and \eq{ymu}. This can be written as $b'+\theta \, b=0$ and  $U_{Yb}\,b'+U_{YY}\,Y'+\theta\;U_Y=0$\,, where $\theta=\nabla_\mu\,u^\mu$ is the expansion, that we introduced in Section \ref{pivo}. Combining these two equations to eliminate $\theta$, we get 
\begin{align} \label{Nc}
b\,Y' =c_b^2\,Y\,b'\,.
\end{align}
Using this result, the sound speed \eq{so} can be easily computed as
(see also \cite{Nicolis:2011cs})
\be
\label{cs}
\begin{split}
c_s^2=\frac{ p_b\, b'+p_Y\,
   Y'}{\rho_b\,b'+\rho_Y\,
   Y'}
%=\frac{(b\;U_{bY}-U_Y)^2-b^2\;U_{bb}\,U_{YY}}{U_{YY}\,(\rho+p)}
=\frac{c_b^4\,Y\,\rho_Y+b\,p_b}{\rho+p}\,,
\end{split}
\ee
 where, as we have been doing through, the subscripts in $U$, $\rho$ and $p$ denote partial derivatives, e.g.\
\be \label{prho}
  \rho_Y=\left. \frac{\partial \rho}{\partial Y}\right |_b=Y\,U_{YY} \,,\quad p_b =\left. \frac{\partial p}{\partial b}\right |_Y=-b\,U_{bb}\,. 
 \ee
Unsurprisingly, \eq{cs} reduces to one of the expressions \eq{speeds} if $b$ or $Y$ are absent from the action. The quantities \eq{Nc}, \eq{cs} and \eq{prho} allow to express generic variations of the pressure and energy density as follows:
\ba \label{genv}
\delta p=p_b\,\delta b+ c_b^2\,\rho_Y\,\delta Y \,,\quad \delta \rho=\rho_Y\,\delta
 Y+c_Y^{-2}\,p_b\,\delta b\,,
 \ea
so clearly, $\delta p\neq c_s^2\,\delta\rho$. The missing ingredient that allows to turn this into an equality is the variation of the entropy density per particle, $\sigma=s/n$, which we can compute using the dictionary of Table \ref{tab:thermo}, obtaining:
\begin{align}
\delta \sigma=\alpha\,\frac{\rho_Y}{b} \left(\frac{\delta Y}{Y}-c_b^2\,\frac{\delta
 b}{b}\right) \, , \qquad \alpha =\begin{cases} - (b/U_Y)^{2}=-\sigma^2 \,,  &
     \text{for } \{\mu , \, s \} \\
1\,,  & \text{for } \{n ,\,
T \} \end{cases}\,, \label{dsigma}
\end{align}
so that
\be \label{dsig}
\delta p=c_s^2\; \delta\rho+\frac{b \, Y}{
 \alpha}\left(c_b^2-c^2_s\right)\;\delta \sigma\,.
\ee
 We recall that in a perfect fluid --for which the entropy and
 particle currents are parallel--if one of them is conserved the
 other is also conserved, which implies that $\sigma'=0$; see Section
 \ref{pivo}. This is precisely what happens in the case at hand, where
 $\mathcal{Y^\mu}$ and $\mathcal{J^\mu}$  are aligned with $u^\mu$;
 see Table \ref{tab:thermo} for their interpretation according to the
 choice of thermodynamic variables.
 Notice also that $\sigma'=0$ does not imply in general  that $\sigma$ is
 vanishing.
It is important to emphasize that these results rely on the thermodynamic interpretation of the EFT action, which allows us to identify the currents of entropy and density. Clearly, the thermodynamic interpretation is also needed  to compute $\sigma$ and its variation \eq{dsigma}. It is also worth stressing that these arguments are valid at all orders in fluctuations, as it is clear from the way we have constructed the variations in \eq{dsigma}.

\section{Propagation of phonons and dynamical stability} \label{prop}

In this section we study the propagation of the internal degrees of
freedom of effective perfect fluids and their relation to energy and
pressure perturbations, as well as the conditions that ensure
dynamical stability. The results of this analysis will be related to
thermodynamic stability, which was discussed in Section
\ref{sect:stab}.  For simplicity, we will consider perfect fluids
living in flat spacetime. Then, the dynamics of linear fluctuations
are given by the {\it Euler} and the {\it continuity } equations (EE
and CE), which read, respectively,\footnote{In this section, both
  $\partial_t$ and the overdots indicate time derivatives. The background enthalpy per unit of volume, denoted here by $\rho+p$, is a constant in Minkowski spacetime.}
\begin{align} \label{Euler}
(\rho+p)\,\partial_t\,v^i+\partial_i\,\delta p & = 0\,,\\
\partial_t\,{\delta\rho}+(\rho+p)\,\partial_i\, v^i & = 0 \label{cont}
\end{align}
and come from the conservation of $T_{\mu\nu}$, see \eq{pfEMT} and
\eq{T4}. To solve them, an extra relation between $\delta p$,
$\delta\rho$ and $\partial_i\, v^i$ needs to be known, which in our
case is provided by the EFT action. 
The expansion  at the second order in the phonon fields of the three
operators of the EFT is given by
\ba
b&=&1+\partial_i\pi^i-\frac{1}{2}\dot
\pi^i\dot\pi^i+\frac{1}{2}(\partial_i\pi^i)^2-\frac{1}{2}(\partial_i\pi^j)(\partial_j\pi^i)
\, ; \\
 Y&=&1+\dot\pi^0+\frac{1}{2}\dot
 \pi^i\dot\pi^i-\dot\pi^i\partial_i\pi^0 \, ;
 \\
 X&=& -1-2\, \dot\pi^0-(\dot\pi^0)^2 - (\partial_i\dot \pi^0)^2 \, .
\ea
\newline

\noindent \textbf{\textit{i)}} Let us start with $U(b)$ and write the scalar fields as
\begin{align} \label{phonons}
\Phi^i(t,x^j) = x^i +\pi^i(t, \vec{x})\,.
\end{align}
In this expression $\pi^i$ represent the {\it phonons} that are
fluctuations around the static background $\Phi^i=x^i$ under the
assumption that $|\partial_i\pi^j|\ll 1$. Indeed, the effective theory
of fluids that we are discussing can be seen as the theory of the
propagation of sound waves in continuous media
\cite{Leutwyler:1996er}. Notice that the fields $\pi^i=\Phi^i-x^i$ are
invariant under the combination of a constant translation
$x^i\rightarrow x^i-c^i$ and an internal shift $\Phi^i\rightarrow
\Phi^i+c^i$. Therefore, the $\pi^i$ propagate on a homogeneous
background and represent the Goldstone bosons of broken translations. 
 It is convenient to split the phonons into a longitudinal component
 and two transverse ones, defined by
\begin{align}
\pi^i=\pi^i_L+\pi^i_T,\quad \pi_L^i= \de_i \pi_L \,,\quad \partial_i\pi^i_T=0. 
\end{align} 
The action expanded at quadratic order reads:
\begin{align} 
S^{(2)}[b]=\frac{1}{2}(\rho+p)\int d^4x\left[\dot\pi_T^i\dot\pi_T^i-
  \dot{\pi_L} \Delta \, \dot{\pi}_L 
  -c_s^2\,(\Delta \pi_L)^2\right]\,,
\end{align}
where $\Delta = \de_i \de_i$ and $c_s^2$ can be read in \eq{speeds}.
The equations of motion (EOMs) are $\ddot\pi^i_T=0$ and 
$\ddot\pi_L-c_s^2 \, \Delta \pi_L=0$. Therefore, the
transverse modes do not propagate--their amplitude simply changes
linearly in time--due to the conservation of vorticity, which can be
traced back to the symmetry $V_s$Diff.
The longitudinal
mode propagates with speed of sound given by \eq{speeds}. The linear
pressure and density perturbations are $\delta p=c_s^2\delta
\rho=c_s^2(\rho+p)\, \Delta \pi_L$; and indeed $c_s^2=dp/d\rho$, as 
can be seen directly from \eq{LT}. 
The linear velocity perturbation is $u^i=-\dot \pi^i$ and the
divergence of the EE \eq{Euler} is nothing but the equation for the 
propagation of $\pi_L$. The CE \eq{cont} is simply an
identity.  Stability requires $\rho+p\geq 0$, in order to avoid
possibly dangerous ghosts, and 
$c_s^2\geq 0$, to avoid exponential growth of $\pi_L$. The first
condition is $U_b \leq 0$. If this is satisfied, the constraint 
$c_s^2 \geq 0$ is
equivalent to the condition for thermodynamic stability $U_{bb}\geq 0$  of \eq{ts1}.
\newline

\noindent \textbf{\textit{ii)}} Let us now consider the irrotational perfect fluid $U(X)$. The only scalar that is present in this case is $\Phi^0$, which we write in a similar fashion as \eq{phonons}: $\Phi^0=\varphi(t)+\pi^0$. The EOM for $\varphi$ is $\partial_t(U_x\dot\varphi)=0$ and implies that $\varphi\propto t$ (except if $\rho\rightarrow 0$, which is a limit we discard). So, we write
\begin{align} \label{fon}
\Phi^0=t+\pi^0\,,
\end{align}
and expand the action assuming that $|\partial\pi^0|\ll 1$. The dynamics of the Goldstone boson of the broken time translation is then governed by the action
\begin{align}
S^{(2)}[X]=\frac{1}{2}(\rho+p)\int d^4x\left[c_s^{-2}(\dot\pi^0)^2-(\partial_i\pi^0)^2\right]\,,
\end{align}
where $c_s^2$ is given in \eq{speeds}. The linear pressure and energy
density perturbations are $\delta p=
-2\,p_X\,\dot\pi^0=c_s^2\delta\rho$ and, again,
$c_s^2=dp/d\rho=p_X/\rho_X$ from \eq{LT}. The velocity perturbation is
$\mathcal{V}^i=-\partial_i\pi^0$. The EOM for $\pi^0$ is
$\ddot\pi^0-c_s^2
\, \Delta \pi^0=0$, which is the CE \eq{cont}, whereas the EE \eq{Euler} is now an identity; just the opposite of what occurs for $U(b)$. Dynamical stability requires $(\rho +p)/c_s^2\geq 0$ and $\rho+p\geq 0$. Thermodynamic stability, see \eq{ts2},  demands precisely the first of these two conditions. \newline

\noindent \textbf{\textit{iii)}} We will now consider $U(Y)$, which
contains both types of scalar fields, $\Phi^i$ and $\Phi^0$. The
appropriate background in this case is given by \eq{phonons} and
\eq{fon}, as it can be checked using the EOMs. In particular, the
conservation of \eq{ymu} implies that $U_Y$ is a constant, which then
requires $\Phi^0\propto t$. The quadratic action for the phonons
around the background is
\begin{align} \label{SY}
S^{(2)}[Y]=\frac{1}{2}(\rho+p)\int
  d^4x\left[\dot\pi_T^i\dot\pi_T^i-\dot{\pi}_L \, \Delta \, \dot\pi_L-c_s^{-2}(\dot\pi^0)^2-2\dot\pi^0 \,
  \Delta \pi_L  \right]\,,
\end{align}
where, once more, $c_s^2$ is given in \eq{speeds}. The EOMs derived
from the above action, besides $\ddot{\pi}_T^i =0$,  are
\be
 \ddot{\pi}_L - \dot\pi^0=0 \, , \qquad  \ddot\pi^0-c_s^2 \Delta \, \dot{\pi}_L=0 \, .
\ee
As in the case $U(b)$, $\ddot\pi_T^i=0$ is a consequence of the symmetry $V_s$Diff.  

 Equating to zero the determinant of the quadratic form that defines
 the Lagrangian of \eq{SY} and going to Fourier space, one finds two
 modes: one with a dispersion relation $\omega^2-c_s^2\,k^2=0$ and a 
 second one with $\omega=0$.
 This second mode is similar to the transverse ones, $\pi^i_T$, which
 also have $\omega=0$. Notice, as well, that once the equation of motion
for $\pi_L$ is solved, the dynamics of $\pi^0$ is given by a {\it single} time integration constant. 

As in the case $U(X)$, the energy density and pressure
perturbations depend on 
$\dot\pi^0$, i.e. $\delta p=c_s^2\delta\rho=c_s^2(\rho+p)\dot\pi^0$,
whereas the velocity perturbation is the same as for $U(b)$, i.e.\
$u^i =-\dot\pi^i$. The EE is the first of the EOMs derived from
\eq{SY} and the CE \eq{cont} is the other one, since $u^i$ depends on
$\pi^i$ while $\delta\rho$ depends on $\pi^0$. 
In this case dynamical stability requires $\rho + p>0$ and $c_s^2 >0$
as it can be checked by imposing that the Hamiltonian density derived
from (\ref{SY}) is positive. When $\rho + p= U_Y>0$ holds,  
thermodynamic stability implies 
that $c_s^2>0$,  see \eq{ts1}. \newline

To summarize the results so far, we have seen that for perfect fluids
$U(b)$, $U(X)$ and $U(Y)$ longitudinal modes --a single mode for
$U(b)$ and $U(X)$ and two for $U(Y)$-- propagate with speeds of sound
$c_s^2=dp/d\rho$, given in \eq{speeds}. This reflects that these
fluids are barotropic. 
We have seen that in these three cases, thermodynamic stability plus the null-energy condition imply positivity of the speed of sound, ensuring dynamical stability.
\newline

\noindent \textbf{\textit{iv)}} Finally, let us focus on $U(b,Y)$. Expanding as before around $\Phi^i=x^i$ and $\Phi^0=t$, we obtain 
\begin{align} 
\label{SbY}
S^{(2)}[b,Y]=\frac{1}{2}\int
  d^4x\Big[(\rho+p) \left( \dot\pi_T^i\dot\pi_T^i -\dot{\pi}_L \,
  \Delta \, \dot\pi_L \right)+\rho_Y(\dot\pi^0)^2-p_b\,(\Delta
  \pi_L)^2-2\,c_b^2\,\rho_Y\,\dot\pi^0 \, \Delta \pi_L         \Big]\,.
\end{align}
As before, $V_s$Diff gives $\ddot\pi^i_T=0$, and the remaining EOMs are 
\be
\ddot\pi^0-c_b^2\,\Delta \dot \pi_L=0 \, , \qquad 
(\rho+p)\ddot \pi_L-\rho_Y\,c_b^2\, \dot\pi^0-p_b\, \Delta \, \pi_L=0
\, ;
\ee
which give
\be\label{eqst}
\dot \pi^0=c_b^2\;\Delta \pi_L +\sigma_0(\vec{x})
\,,\quad \ddot \pi_L-c_s^2\;\Delta\pi_L=\frac{\rho_Y\;c_b^2}{\rho+p}\; \sigma_0(\vec{x})
\ee 
where $\sigma_0(\vec{x})$ is a generic time-independent function,
fixed by initial conditions.  
The situation is analogous to the case $U(Y)$ with two modes, one that
propagates with velocity $c_s^2$ given by \eq{cs}, and a second one with $\omega=0$. The reason is the lack of a quadratic term in \eq{SbY} containing only spatial derivatives of $\pi^0$.

The  Hamiltonian  density  ${\cal H}$ derived from (\ref{SbY}) is given by
${\cal H} = P_0 \, \dot{\pi}^0 + P_L \, \dot{\pi}_L + \vec{P} \cdot
\dot{\vec{\pi}}_T - {\cal L}$, where ${\cal L}$ is the Lagrangian
density and  $P_0$, $P_L$, $\vec{P}$  are the momenta conjugate to
$\pi^0$, $\pi_L$ and $\vec{\pi}_T$ 
respectively. As a result 
\be
{\cal H} =
\frac{1}{2} \left[ 
   (p+\rho )\left(\dot{\vec{\pi}}_T \cdot \dot{\vec{\pi}}_T+\vec{\nabla}\dot{\pi}_L \cdot \vec{\nabla}\dot{\pi}_L\right) + U_{YY}\, \left(\dot{\pi}^0\right)^2
    - U_{bb} \, \left(\Delta \pi _L\right)^2 \right] \,,
\ee
Thus ${\cal H}$ is positive definite when $\rho + p = U_Y - U_b>0$, $U_{bb} <0$  and
$U_{YY} >0$.
Notice that the previous three conditions also ensure
that $c_s^2>0$,  see \eq{cs}.
The entropy perturbation is nonvanishing. Indeed at linear order, by using the dictionary entry
for $U(b, \, Y)$ relative to $(n, \, T)$, we have that 
\be
\delta \sigma \equiv
\rho_Y\;\left(\dot \pi^0-c_b^2\;\Delta \pi_L
\right)=\rho_Y\;\sigma_0(\vec{x})  \, . 
\ee
It is straightforward to see from \eq{eqst} that  although $\delta \sigma \neq 0$, it  is constant
in time
\be
\de_t ( \delta \sigma)=0 \, ;
\ee
as expected for an adiabatic fluid. The conclusion is
the same when  the dictionary entry for $U(b, \, Y)$ relative to $(\mu, \, s)$ is used. 
The analysis of Section \ref{sect:stab}
showed that thermodynamic stability requires $U_{bb}\leq0$ and 
$\,U_{YY}\geq 0$. As in the previous cases,  thermodynamic  stability plus the
null-energy condition imply dynamical stability.
 Notice that  in the
cases where an unspecified function appears in the dictionary,  thermodynamic stability implies 
additional constraints on $f$ that cannot be obtained requiring dynamical stability of the phonons. However, in the examples where $f$ is known, such as those of Section \ref{sect:app}, those 
additional constraints are automatically satisfied.

\section{Summary and conclusions}    
 
 \label{sect:conc}

We have studied thermodynamic interpretation of
the EFT of perfect fluids, obtaining a dictionary summarized in Table
\ref{tab:thermo}, completing and extending  the results  of~\cite{Dubovsky:2005xd,Dubovsky:2011sj}. 
We have established the correspondence  between the fundamental thermodynamic variables needed to describe simple 
systems and the EFT operators that configure the four types of perfect fluids that are allowed in the theory. Each entry of the dictionary (described by Table \ref{tab:thermo}) corresponds to a specific operator content in the EFT. The interpretation of the EFT master function $U$--the Lagrangian--as a  thermodynamic potential is determined by the EMT. For the effective perfect fluids depending on a single scalar operator, the master function  $U$ represents either the energy density  (for $U(b)$) or the pressure (for  $U(Y)$ and $U(X)$). 
For these cases, the thermodynamic potentials ${\cal F}$ and ${\cal I}$ appears as  specific limits  of the free function $f$ of Table \ref{tab:thermo}.
 For the Lagrangian $U(b,\,Y)$, the master function corresponds to  a Legendre transformation  of the energy density (or the pressure) to another thermodynamic potential, such as 
the free energy $\mathcal{F}=\rho-T\,s$ or to the potential $\mathcal{I}=\rho-\mu\,n$. 

A full thermodynamic correspondence,  allowing to identify simultaneously an independent and conserved particle (or charge) number current and a conserved entropy density current is only possible for perfect fluids described by the Lagrangian $U(b,Y)$.

This action 
 is invariant under the internal symmetry group $V_s$Diff $\times\, T_s$ of spatial volume preserving diffeomorphisms and time redefinitions that depend on the spatial fields. 
 
The fact that $U(b,Y)$ is the only perfect fluid Lagrangian that is chosen by a (continuous) symmetry (assuming four scalars), highlights it even more as the most complete effective description of perfect fluids, see also \cite{Dubovsky:2011sj}. The other perfect fluids, indicated as $U(b)$, $U(Y)$ and $U(X)$ in Table \ref{tab:thermo}, can also be given thermodynamic interpretations, but only in limits where a single thermodynamic variable is sufficient for the description. Clearly, this is because two independent operators ($b$ and $Y$) are needed for a nondegenerate matching with two independent thermodynamic variables; and also to describe two independent thermodynamic conserved currents.  We have illustrated the use of the thermodynamic dictionary with a few well-known cases of perfect fluids, such as e.g.\ Bose-Einstein and Fermi-Dirac gases in the deeply relativistic limit and a van der Waals gas. 

We have argued that internal shift symmetries are a necessary
condition for a thermodynamic description by studying the Lagrangian
for a single scalar with an explicitly broken shift symmetry:
$U(X,\phi)$, whose EMT has the form of a perfect fluid, showing that in
such a case the basic thermodynamic relations are incompatible with
the  equation of motion of the field.

We have also studied the propagation of linearized sound waves in flat
spacetime and how the Euler and the continuity equations describe the
dynamics of Goldstone bosons for each kind of perfect fluid. This
analysis leads to the conclusion  that thermodynamic stability plus
the null-energy condition, $\rho+p\geq 0$, ensure dynamical
stability. This  holds true for the four possible types of effective
perfect fluids. 
The same
analysis shows that  the fluid described by $U(b,Y)$, being in
general nonbarotropic, can support nonvanishing entropy per particle
perturbations but is  nonetheless  adiabatic.

\section*{Acknowledgements}
We thank B.\ Bellazzini, L.\ Hui, I.\ Sawicki and S.\ Sibiryakov for
useful discussions. L.P. thanks S. Ciuchi for interesting discussions
and suggestions. The work of G.B.\ is funded by the European Union's
Horizon 2020 research and innovation programme under the Marie
Sk\l{}odowska-Curie grant agreement number 656794. G.B.\ thanks  the
CERN Theoretical Physics Department for hospitality while this work
was done. L.P.\ thanks  the Institute de Physique Th\'eorique IPhT
CEA-Saclay for hospitality. We also thank the Galileo Galilei
Institute for Theoretical Physics for hospitality and the Istituto
Nazionale di Fisica Nucleare (INFN) for support during the completion of this work.

\begin{appendix}

\section{Appendix: A systematic approach to the thermodynamic dictionary}
\label{app}
Any simple thermodynamic system is described by at most  six variables: 
$\zeta=\{s,\,T,\,n,\,\mu,\,\rho,\,p\}$, of which only two are
independent. For the EFT of perfect fluids, $\rho$ and $p$ are given
as functions of at most two operators among ${\cal
  O}=\{b,\,Y,\,X\}$. Let us choose two independent variables (other
than $\rho$ and $p$) from the list $\zeta$ and denote them $z_1$ and
$z_2$. These can be, for instance, $\{n,s\}$ or $\{\mu,T\}$, see Table
\ref{tab:Leg}. Since all the other four variables contained in
$\zeta$ are, by assumption, functions of $z_1$ and $z_2$, we can formally write
 \be
\zeta(z_1,z_2)=\underbrace{\{z_1,\,z_2\}}_{\rm Indep.\,var.}\times \underbrace{\left\{{\cal Z}_{1}(z_1,\,z_2)\,,{\cal Z}_{2} (z_1,\,z_2)\right\}}_{\rm Dependent\;variables}\times \underbrace{\left\{
 \rho\left[{\cal O}(z_1,\,z_2)\right],\,
 p\left[{\cal O}(z_1,\,z_2)\right]\right\}}_{\rm Variables\; from\; the\; EMT}\,,
 \ee
where we have explicitly taken into account that the operators $\cal O$ must be functions of $z_1$ and $z_2$, and we have denoted by ${\cal Z}_{1}$ and ${\cal Z}_{2}$ the two (dependent) variables that are not $\rho$ and $p$. The thermodynamic relations \eq{termo1}, \eq{termo2} and \eq{termo3}  constrain the dependence of ${\cal Z}_{1,2}(z_1,\,z_2) $ and ${\cal O}(z_1,\,z_2)$ on the chosen independent variables $z_1$ and $z_2$. The thermodynamic dictionary for the EFT of perfect fluids can be derived, requiring that:
\begin{itemize}
\item
It has to be valid for any master function $U$.
\item
The operators ${\cal O}$ are independent of 
$U$ and its  derivatives.
\end{itemize}
Then, any thermodynamic constraint involving the derivatives $U^{(n)}$ must hold irrespectively of the form of $U^{(n)}$. Moreover, from the Euler relation (\ref{termo2}), we see that the dependent variables ${\cal Z}_{1,2}(z_1,\,z_2) $  will depend on first derivatives of $U$. One can easily keep track  of the number of derivatives entering in each quantity using a counting parameter $\epsilon$, such that
 \begin{align}
 \begin{aligned} \label{ctd}
U^{(n )}({\cal O}) & \to \epsilon^{n}\;U^{(n)}({\cal O}) \,,
 \\ 
 {\cal Z}_{1,2}^{(n_1,\,n_2)}(z_1,\,z_2) & \to \epsilon^{1+n_1+n_2}\;{\cal Z}_{1,2}^{(n_1,\,n_2)}(z_1,\,z_2)\,,\\
 {\cal O}^{(n_1,\,n_2)}(z_1,\,z_2) & \to \epsilon^{0}\;{\cal O}
 ^{(n_1,\,n_2)}(z_1,\,z_2)\, . 
\end{aligned}
\end{align}
As an example, let us take the case $U(b, \, Y)$ with $\{z_1,\,z_2\}=\{n,\, T\}$
as independent variables. From \eq{UbY} we have that  
$\rho = Y \, U_Y - U$ and  $p=  U - b \, U_b$. Then, $\{{\cal Z}_{1},\,{\cal
  Z}_{2}\}=\{s(n,\, T),\,\mu(n,\, T)\}$ and ${\cal O}=(Y,\,b)$ are also  functions  of $n$ and $T$.
The Euler relation (\ref{termo2}) reads
\be
\label{eu}
\epsilon\;(Y \, U_Y- b \, U_b) =\epsilon \left[T\;s(n,T) +n\;\mu(n,T) \right]
\ee
consistently with \eq{ctd}. Solving for
 $s(n,T)$ and inserting it in (\ref{termo1}), or equivalently in (\ref{termo3}), we get
\be
\begin{split}
&T\,\left \{
 \epsilon^2 \left[b \left(-b_n \,U_{b^2}-Y_n\,
   U_{b Y}\right)-n \,\mu _n\right]+ \epsilon\,
   Y_n \,U_Y
\right\}dn+\\
 &
 \left \{ \epsilon^2\, T\left[b \,\left(-b_T \,U_{b^2}-Y_T\,
   U_{b Y}\right)-n\, \mu _T \right]+ \epsilon
   \left({b \,U_b}+{n\,\mu \,
   -Y\, U_Y}+T\,Y_T \,U_Y\right)
 \right \}\;dT=0 \, .
\end{split} 
\ee
The coefficients of $dn$ and  $dT$ must vanish independently, for all $U$ and order by order in $\epsilon$. At order $\epsilon$ we get $\epsilon\,Y_n=0$ and $\epsilon(b\,U_b+n\,\mu-Y\,U_Y+T\, Y_T\,U_Y)=0$. Differentiating these equations with respect to $n$ and $T$ we obtain $\epsilon^2\, Y_{nn} =\epsilon^2\,Y_{n T}=0$ and
 \begin{align}\nonumber
\epsilon^2 \,\mu _n\,n^2 &= \epsilon^2\, n \,b_n
   \left[\left(Y-T \,Y_T\right) U_{b
   Y}-b \,U_{b^2}\right]+\epsilon\,
   \left[\left(b-n\, b_n\right)
   U_b+\left(T \,Y_T-Y\right)
   U_Y\right] \,,\\
   \epsilon^2\, n \mu _T & = \epsilon^2 \left[-b\,
   b_T \,U_{b^2}+\left(Y \,b_T-\left(T\,
   b_T+b\right) Y_T\right) U_{b
   Y}+Y_T \,\left(Y-T\, Y_T\right)
   U_{Y^2}\right]-\epsilon\,
   \left(b_T \,U_b+T\, Y_{T^2}\,
   U_Y\right)\,. \nonumber
 \end{align}
New  terms of order $\epsilon$ appear and  they have to vanish,
leading to $\left(b-n\, b_n\right)
   U_b+\left(T \,Y_T-Y\right)
   U_Y=0$ and $b_T \,U_b+T\, Y_{T^2}\,
   U_Y =0 $. The solution of these last equations (independent from the form of $U$)  is
$ b=n\;b_n$, $Y=T\;Y_T$, $b_T=Y_{TT}=0$ and, therefore,
 \be \label{solx}
 b=n,\quad Y=T \, ,\quad \mu=-U_b,\quad s=U_Y  \, .
 \ee
One can now check that all the thermodynamic relations are satisfied at all orders in $\epsilon$. The solution \eq{solx}  gives the entry of Table \ref{tab:thermo} relative to $U(b, \, Y)$ with $(n, \, T)$ as independent variables in our dictionary. All the other entries can be derived in the same way. 

\end{appendix}

\bibliographystyle{hunsrt}  
  
\bibliography{fluidbiblio}

\begin{thebibliography}{10}

\bibitem{Andersson:2006nr}
N.~Andersson and G.~L. Comer.
\newblock {Relativistic fluid dynamics: Physics for many different scales}.
\newblock {\em Living Rev. Rel.}, 10:1, 2007, gr-qc/0605010.

\bibitem{Leutwyler:1993gf}
H.~Leutwyler.
\newblock {Nonrelativistic effective Lagrangians}.
\newblock {\em Phys. Rev.}, D49:3033--3043, 1994, hep-ph/9311264.

\bibitem{Leutwyler:1996er}
H.~Leutwyler.
\newblock {Phonons as goldstone bosons}.
\newblock {\em Helv. Phys. Acta}, 70:275--286, 1997, hep-ph/9609466.

\bibitem{Dubovsky:2005xd}
S.~Dubovsky, T.~Gregoire, A.~Nicolis, and R.~Rattazzi.
\newblock {Null energy condition and superluminal propagation}.
\newblock {\em JHEP}, 03:025, 2006, hep-th/0512260.

\bibitem{Dubovsky:2011sj}
Sergei Dubovsky, Lam Hui, Alberto Nicolis, and Dam~Thanh Son.
\newblock {Effective field theory for hydrodynamics: thermodynamics, and the
  derivative expansion}.
\newblock {\em Phys. Rev.}, D85:085029, 2012, 1107.0731.

\bibitem{Nicolis:2015sra}
Alberto Nicolis, Riccardo Penco, Federico Piazza, and Riccardo Rattazzi.
\newblock {Zoology of condensed matter: Framids, ordinary stuff, extra-ordinary
  stuff}.
\newblock {\em JHEP}, 06:155, 2015, 1501.03845.

\bibitem{Ballesteros:2016gwc}
Guillermo Ballesteros, Denis Comelli, and Luigi Pilo.
\newblock {Massive and modified gravity as self-gravitating media}.
\newblock 2016, arXiv:1603.02956.

\bibitem{Nicolis:2013lma}
Alberto Nicolis, Riccardo Penco, and Rachel~A. Rosen.
\newblock {Relativistic Fluids, Superfluids, Solids and Supersolids from a
  Coset Construction}.
\newblock {\em Phys. Rev.}, D89(4):045002, 2014, 1307.0517.

\bibitem{Ballesteros:2014sxa}
Guillermo Ballesteros.
\newblock {The effective theory of fluids at NLO and implications for dark
  energy}.
\newblock {\em JCAP}, 1503(03):001, 2015, 1410.2793.

\bibitem{Matarrese:1984zw}
S.~Matarrese.
\newblock {On the Classical and Quantum Irrotational Motions of a Relativistic
  Perfect Fluid. 1. Classical Theory}.
\newblock {\em Proc. Roy. Soc. Lond.}, A401:53--66, 1985.

\bibitem{Callen}
H.~Callen.
\newblock Thermodynamics and an introduction to thermostatistics, 2nd edition.
\newblock {\em Wiley}, 1985.

\bibitem{rezzolla2013relativistic}
L.~Rezzolla and O.~Zanotti.
\newblock {\em Relativistic Hydrodynamics}.
\newblock OUP Oxford, 2013.

\bibitem{Carter:1987qr}
Brandon Carter.
\newblock {Covariant Theory of Conductivity in Ideal Fluid or Solid Media}.
\newblock {\em Lect. Notes Math.}, 1385:1--64, 1989.

\bibitem{Son:2005ak}
D.~T. Son.
\newblock {Effective Lagrangian and topological interactions in supersolids}.
\newblock {\em Phys. Rev. Lett.}, 94:175301, 2005, cond-mat/0501658.

\bibitem{Endlich:2010hf}
Solomon Endlich, Alberto Nicolis, Riccardo Rattazzi, and Junpu Wang.
\newblock {The Quantum mechanics of perfect fluids}.
\newblock {\em JHEP}, 04:102, 2011, 1011.6396.

\bibitem{Nicolis:2011cs}
Alberto Nicolis.
\newblock {Low-energy effective field theory for finite-temperature
  relativistic superfluids}.
\newblock 2011, 1108.2513.

\bibitem{Endlich:2012pz}
Solomon Endlich, Alberto Nicolis, and Junpu Wang.
\newblock {Solid Inflation}.
\newblock {\em JCAP}, 1310:011, 2013, 1210.0569.

\bibitem{Ballesteros:2012kv}
Guillermo Ballesteros and Brando Bellazzini.
\newblock {Effective perfect fluids in cosmology}.
\newblock {\em JCAP}, 1304:001, 2013, 1210.1561.

\bibitem{Bhattacharya:2012zx}
Jyotirmoy Bhattacharya, Sayantani Bhattacharyya, and Mukund Rangamani.
\newblock {Non-dissipative hydrodynamics: Effective actions versus entropy
  current}.
\newblock {\em JHEP}, 02:153, 2013, 1211.1020.

\bibitem{Hoyos:2012dh}
Carlos Hoyos, Boaz Keren-Zur, and Yaron Oz.
\newblock {Supersymmetric sound in fluids}.
\newblock {\em JHEP}, 11:152, 2012, 1206.2958.

\bibitem{Ballesteros:2013nwa}
Guillermo Ballesteros, Brando Bellazzini, and Lorenzo Mercolli.
\newblock {The effective field theory of multi-component fluids}.
\newblock {\em JCAP}, 1405:007, 2014, 1312.2957.

\bibitem{Delacretaz:2014jka}
Luca~V. Delacr\'etaz, Alberto Nicolis, Riccardo Penco, and Rachel~A. Rosen.
\newblock {Wess-Zumino Terms for Relativistic Fluids, Superfluids, Solids, and
  Supersolids}.
\newblock {\em Phys. Rev. Lett.}, 114(9):091601, 2015, 1403.6509.

\bibitem{Gripaios:2014yha}
Ben Gripaios and Dave Sutherland.
\newblock {Quantum Field Theory of Fluids}.
\newblock {\em Phys. Rev. Lett.}, 114(7):071601, 2015, 1406.4422.

\bibitem{ArkaniHamed:2002sp}
Nima Arkani-Hamed, Howard Georgi, and Matthew~D. Schwartz.
\newblock {Effective field theory for massive gravitons and gravity in theory
  space}.
\newblock {\em Annals Phys.}, 305:96--118, 2003, hep-th/0210184.

\bibitem{Dubovsky:2004sg}
S.~L. Dubovsky.
\newblock {Phases of massive gravity}.
\newblock {\em JHEP}, 10:076, 2004, hep-th/0409124.

\bibitem{Rubakov:2008nh}
V.~A. Rubakov and P.~G. Tinyakov.
\newblock {Infrared-modified gravities and massive gravitons}.
\newblock {\em Phys. Usp.}, 51:759--792, 2008, 0802.4379.

\bibitem{Comer:1993zfa}
G.~L. Comer and D.~Langlois.
\newblock {Hamiltonian formulation for multi-constituent relativistic perfect
  fluids}.
\newblock {\em Class. Quant. Grav.}, 10:2317--2327, 1993.

\bibitem{Comer:1994tw}
G.~L. Comer and D.~Langlois.
\newblock {Hamiltonian formulation for relativistic superfluids}.
\newblock {\em Class. Quant. Grav.}, 11:709--721, 1994.

\bibitem{Comer:2011ss}
G.~L. Comer, Patrick Peter, and N.~Andersson.
\newblock {Multi-fluid cosmology: An illustration of fundamental principles}.
\newblock {\em Phys. Rev.}, D85:103006, 2012, 1111.5043.

\bibitem{Blas:2012vn}
Diego Blas, Mikhail~M. Ivanov, and Sergey Sibiryakov.
\newblock {Testing Lorentz invariance of dark matter}.
\newblock {\em JCAP}, 1210:057, 2012, 1209.0464.

\bibitem{Pourtsidou:2013nha}
A.~Pourtsidou, C.~Skordis, and E.~J. Copeland.
\newblock {Models of dark matter coupled to dark energy}.
\newblock {\em Phys. Rev.}, D88(8):083505, 2013, 1307.0458.

\bibitem{Berezhiani:2015pia}
Lasha Berezhiani and Justin Khoury.
\newblock {Dark Matter Superfluidity and Galactic Dynamics}.
\newblock {\em Phys. Lett.}, B753:639--643, 2016, 1506.07877.

\bibitem{Kopp:2016mhm}
Michael Kopp, Constantinos Skordis, and Dan~B. Thomas.
\newblock {An extensive investigation of the Generalised Dark Matter model}.
\newblock 2016, 1605.00649.

\bibitem{Kang:2015uha}
Jonghee Kang and Alberto Nicolis.
\newblock {Platonic solids back in the sky: Icosahedral inflation}.
\newblock {\em JCAP}, 1603(03):050, 2016, 1509.02942.

\bibitem{khalatnikov1982relativistic}
I.~M. Khalatnikov and V.~V. Lebedev.
\newblock Relativistic hydrodynamics of a superfluid liquid.
\newblock {\em Physics Letters A}, 91(2):70--72, 1982.

\bibitem{PhysRevD.45.4536}
B.~Carter and I.~M. Khalatnikov.
\newblock Equivalence of convective and potential variational derivations of
  covariant superfluid dynamics.
\newblock {\em Phys. Rev. D}, 45:4536--4544, Jun 1992.

\bibitem{ArmendarizPicon:1999rj}
C.~Armendariz-Picon, T.~Damour, and Viatcheslav~F. Mukhanov.
\newblock {k - inflation}.
\newblock {\em Phys. Lett.}, B458:209--218, 1999, hep-th/9904075.

\bibitem{ArmendarizPicon:2000dh}
C.~Armendariz-Picon, Viatcheslav~F. Mukhanov, and Paul~J. Steinhardt.
\newblock {A Dynamical solution to the problem of a small cosmological constant
  and late time cosmic acceleration}.
\newblock {\em Phys. Rev. Lett.}, 85:4438--4441, 2000, astro-ph/0004134.

\bibitem{Deffayet:2010qz}
Cedric Deffayet, Oriol Pujolas, Ignacy Sawicki, and Alexander Vikman.
\newblock {Imperfect Dark Energy from Kinetic Gravity Braiding}.
\newblock {\em JCAP}, 1010:026, 2010, 1008.0048.

\bibitem{Pujolas:2011he}
Oriol Pujolas, Ignacy Sawicki, and Alexander Vikman.
\newblock {The Imperfect Fluid behind Kinetic Gravity Braiding}.
\newblock {\em JHEP}, 11:156, 2011.

\bibitem{askey1982}
L~Lewine.
\newblock {\em Polylogarithms and associated functions"}.
\newblock Elsevier Science Ltd, 1982.

\bibitem{LandauStat}
L.D. Landau and E.M. Lifishitz.
\newblock {\em Statistical Physics, part I}.
\newblock Elsevier, 1980.

\end{thebibliography}
  
\end{document}